\documentclass[sn-mathphys-num]{sn-jnl}

\usepackage{bm}
\usepackage{graphicx}%
\usepackage{subcaption}
\usepackage{multirow}%
\usepackage{amsmath,amssymb,amsfonts}%
\usepackage{amsthm}%
\usepackage{mathrsfs}%
\usepackage[title]{appendix}%
\usepackage{xcolor}%
\usepackage{textcomp}%
\usepackage{manyfoot}%
\usepackage{booktabs}%
\usepackage{algorithm}%
\usepackage{algorithmicx}%
\usepackage{algpseudocode}%
\usepackage{listings}%
\usepackage{hyperref}
\usepackage{longtable}   
\usepackage{booktabs}    
\usepackage{float}

\theoremstyle{thmstyleone}%
\theoremstyle{thmstyletwo}%

\theoremstyle{thmstylethree}%

\begin{document}

\title[Article Title]{Hardware-Aware Quantum Kernel Design Based on Graph Neural Networks}


\author[1]{\fnm{Fanxu} \sur{Meng}}

\author[2,3,4]{\fnm{Yuxiang} \sur{Liu}}

\author[3,4,5]{\fnm{Lu} \sur{Wang}}

\author[4,5]{\fnm{Sixuan} \sur{Li}}

\author*[3,4,5]{\fnm{Xutao} \sur{Yu}}\email{yuxutao@seu.edu.cn}

\author*[2,3,4]{\fnm{Zaichen} \sur{Zhang}}\email{zczhang@seu.edu.cn}


\affil[1]{\orgdiv{College of Artificial Intelligence}, \orgname{Nanjing Tech University}, \orgaddress{\street{Puzhu South Road}, \city{Nanjing}, \postcode{211800}, \country{China}}}

\affil[2]{\orgdiv{National Mobile Communications Research Laboratory}, \orgname{Southeast University}, \orgaddress{\street{Southeast University Road}, \city{Nanjing}, \postcode{210096}, \country{China}}}

\affil[3]{\orgname{Purple Mountain Lab}, \orgaddress{\street{Mozhou East Road}, \city{Nanjing}, \postcode{211111}, \country{China}}}

\affil[4]{\orgdiv{Frontiers Science Center for Mobile Information Communication and Security}, \orgname{Southeast University}, \orgaddress{\street{Southeast University Road}, \city{Nanjing}, \postcode{210096}, \country{China}}}

\affil[5]{\orgdiv{State Key Laboratory of Millimeter Waves}, \orgname{Southeast University}, \orgaddress{\street{Southeast University Road}, \city{Nanjing}, \postcode{210096}, \country{China}}}


\abstract{
Quantum kernels hold significant promise for achieving computational advantages in quantum machine learning (QML), yet their effectiveness critically depends on the design of expressive and hardware-compatible feature maps, a challenge that is particularly pronounced on Noisy Intermediate-Scale Quantum (NISQ) devices with limited qubits, gate errors, and restricted connectivity. In this work, we propose a hardware-aware framework for automated quantum kernel design that integrates quantum device characteristics with learning-based evaluation. Specifically, candidate quantum circuits explored within the hardware-aware circuit space are represented as directed acyclic graphs (DAGs) encoding hardware-specific information such as gate operations, qubit interactions, and noise properties, while a dual graph neural network (GNN) predictor is employed to estimate key surrogate metrics, including probability of successful trials (PST) and kernel-target alignment (KTA), enabling efficient and accurate assessment of circuit fidelity and kernel performance to facilitate the identification of task-specific quantum kernels. Furthermore, feature selection is incorporated to reduce input dimensionality and ensure compatibility with near-term devices. Extensive experiments on multiple benchmark datasets, including Credit Card (CC), MNIST-5, and FMNIST-4, demonstrate that our method consistently outperforms existing baselines in classification accuracy, effectively balancing hardware constraints and model expressivity under realistic noise conditions. These results highlight the potential of combining hardware-aware design with deep learning techniques to advance practical quantum kernel methods and facilitate their deployment on near-term quantum hardware.


%
}






\maketitle

\section{Introduction}\label{sec1}
Quantum computing presents a compelling prospect for revolutionizing machine learning tasks by exploiting the fundamental properties of quantum mechanics, such as superposition, entanglement, and high-dimensional Hilbert spaces~\cite{1,2}. Quantum Machine Learning (QML), as a prominent class of quantum algorithms, has demonstrated practical applicability in diverse real-world tasks, including classification~\cite{16,17,18,anagolum2024elivagar,wang2022quantumnas}, generative modeling~\cite{19,20,21,22}, and learning physical systems~\cite{23,24}, particularly in the Noisy Intermediate-Scale Quantum (NISQ) era~\cite{15}.
Among various QML approaches, quantum kernel methods have attracted increasing attention due to their solid theoretical foundation and interpretability~\cite{3,4,5}. By embedding classical data into quantum states through parameterized quantum circuits, quantum kernels induce implicit feature maps capable of capturing complex nonlinear relationships beyond classical means.

Despite recent demonstrations of quantum kernel learning on real quantum devices, achieving practical quantum advantage remains challenging, particularly under the constraints of Noisy Intermediate-Scale Quantum (NISQ) devices~\cite{6,7,8}. Inappropriate quantum circuit designs can lead to vanishing kernel similarities, reduced expressivity, and degraded generalization~\cite{9,10}, thereby limiting the overall performance of quantum kernels. These issues are further exacerbated in high-dimensional settings or when circuit depth and qubit counts exceed the noise tolerance of near-term hardware.
To address these challenges, prior works have explored both circuit-aware and data-aware optimization strategies. Some approaches focus on tuning variational parameters within fixed circuit architectures~\cite{100,26,27}, while others formulate quantum kernel design as a joint optimization over circuit and parameters, typically addressed using evolutionary or machine learning-based approaches \cite{12,13,14, lei2024neural}. However, these methods not only fail to account for hardware-specific constraints, such as qubit connectivity and gate noise, but also incur prohibitive computational costs when exploring large circuit design spaces.

Motivated by these limitations, there remains a critical need for approaches that can efficiently identify high-quality and task-specific quantum kernel while explicitly incorporating hardware-specific constraints.
To this end, we propose a hardware-aware quantum kernel design framework based on graph neural networks. Rather than formulating quantum kernel design as a joint optimization over circuit structures and parameters, our approach decouples the process by first generating hardware-aware candidate circuits and subsequently evaluating their performance using graph neural networks (GNNs). By exploiting the graph structure of quantum circuits, the GNNs predict two surrogate metrics: the probability of successful trials (PST), which captures circuit fidelity under hardware noise, and kernel-target alignment (KTA), which correlates with downstream classification accuracy.

The proposed framework comprises four stages: generation of hardware-aware candidate circuits based on device topology and native gate sets, filtering via PST prediction to eliminate low-fidelity circuits, ranking via KTA prediction to identify high-performing kernels, and subsequent parameter optimization of the top candidates. Furthermore, to further enhance scalability, we incorporate feature selection to reduce input dimensionality and ensure compatibility with NISQ devices. Crucially, by integrating hardware information and task-dependent performance estimation within a unified learning framework, the proposed framework enables efficient identification of quantum kernels that are both hardware-compatible and task-relevant. Extensive simulations on IBM quantum devices demonstrate that our approach consistently outperforms state-of-the-art baselines across diverse classification tasks and hardware configurations.

Our method offers several distinctive advantages that collectively enable efficient quantum kernel design for NISQ devices:
\begin{itemize}
	\item[$\bullet$] \textbf{Hardware-awareness:} Our methodology explicitly incorporates hardware topology and noise characteristics during both circuit generation and performance prediction, enabling the identification of robust and high-quality quantum circuits across different hardware backends.
	\item[$\bullet$]  \textbf{Dual-GNN-based performance prediction:} By employing two graph neural networks (GNN-1 and GNN-2), the proposed method efficiently predicts PST and kernel-target alignment (KTA) as a proxy for classification performance, achieving speedups of over 380× and 58,000× compared to direct simulation
	\item[$\bullet$]  \textbf{Hardware-aware candidate filtering:} By filtering out low-fidelity circuits early in the pipeline, our framework mitigates the impact of gate errors and decoherence, thereby improving the robustness and reliability of quantum kernel performance on noisy devices.
	\item[$\bullet$] \textbf{Efficient support for high-dimensional input:} The integration of feature selection techniques (e.g., mRMR) reduces input dimensionality while preserving discriminative information, thereby mitigating kernel concentration and enabling practical deployment on NISQ devices.
	\item[$\bullet$] \textbf{Superior classification accuracy:} Extensive evaluations on multiple benchmark tasks, including Credit Card (CC), MNIST-5, and FMNIST-4, as well as across diverse IBM quantum devices, demonstrate that the proposed methodology consistently outperforms existing state-of-the-art methods.
\end{itemize}

The remainder of this paper is organized as follows. Section~\ref{Background} reviews the necessary background on quantum computing, kernel methods, and graph neural networks. Section~\ref{Method} presents the proposed framework. Section~\ref{Evaluation} reports numerical results and comparative evaluations. Finally, Section~\ref{Conclusion} concludes the paper and outlines future research directions.

\section{Background}
\label{Background}
\subsection{Quantum Computing}
A quantum bit (qubit) can exist in a linear superposition of the two computational basis states, $\left | 0  \right \rangle $ and $\left | 1  \right \rangle $,  in contrast to a classical bit that can only be in a definite state of either 0 or 1. The state of a single qubit is generally represented as $\left | \psi  \right \rangle =\alpha \left | 0  \right \rangle + \beta \left | 1  \right \rangle $, where $\alpha $ and $\beta $ are complex amplitudes satisfying the normalization condition $\left | \alpha  \right | ^{2} + \left | \beta   \right | ^{2}=1$. In multi-qubit systems, the overall quantum state resides in a tensor-product Hilbert space, allowing for entangled states that cannot be factorized into products of single-qubit states. Quantum gate, as a unitary matrix, transforms the quantum state to another and preserves the norm of the quantum state. Any unitary can be decomposed into a set of basis gates. The basis gate sets explored in this study are $S=\left \{ R_{x}, R_{z}, CZ, I \right \} $ and $S^{\ast } =\left \{ R_{z}, X, CNOT, I \right \}$, which consist primarily of single-qubit and two-qubit gates. More detailed matrix representations can be found in Appendix \ref{Appendix_Add}.

\subsection{Mechanism of quantum kernel}
\label{Mechanism_of_quantum_kernel}
Kernel methods provide a powerful theoretical framework for performing nonlinear and nonparametric learning tasks, owing to their generality and strong interpretability \cite{mohri2018foundations,aronszajn1950theory,cristianini2001kernel}. Assume that both the training and test samples are drawn from the same input-output space $\mathcal{X} \times \mathcal{Y} $. The training dataset is denoted as $\mathcal{D}=\left \{ \bm{x}_{i},y_{i}    \right \} _{i=1}^{n}\subset   \mathcal{X} \times \mathcal{Y} $, where $\bm{x}_{i}$ represents the input features of the $i$-th sample and $y_{i}$ denotes the corresponding label. The fundamental idea of traditional kernel methods is to map input data $\bm{x}_{i} $ from $\mathbb{R}^{d} $ to a higher-dimensional feature space $\mathbb{R}^{q} $, i.e., $\phi \left ( \cdot  \right ) :\mathbb{R} ^{d} \to \mathbb{R} ^{q}$ with $q\gg d$, such that data that are not linearly separable in the original space may become linearly separable in the transformed space. However, when $q$ is large, explicitly computing the mapped feature vectors $\phi \left ( \bm{x}_{i}   \right ) $ becomes computationally expensive. To address this, kernel methods employ the kernel trick, which enables implicit computation of inner products in the feature space through a kernel function $\kappa(\bm{x}_i, \bm{x}_j)$. This leads to the construction of a kernel matrix $K \in \mathbb{R}^{n \times n}$ with entries defined as
\begin{equation}
\left [ K \right ] _{i,j} = \kappa(\bm{x}_i, \bm{x}_j), \quad \forall i,j \in [n],
\end{equation}
where $\kappa(\bm{x}_i, \bm{x}_j)$ corresponds to the inner product $\langle \phi(\bm{x}_i), \phi(\bm{x}_j) \rangle$ if an explicit feature map exists.
Using the kernel matrix, the learning task reduces to finding a hypothesis function of the form $h\left ( \bm{x}_{i}  \right ) =\left \langle \bm{w}^{\ast }  ,\phi \left ( \bm{x}_{i} \right ) \right \rangle $. The optimal parameter $\bm{w}^{\ast }$ is obtained by minimizing the following loss function:
\begin{equation}
\mathcal{L}=\lambda \left \langle \bm{w}, \bm{w} \right \rangle + \sum_{i=1}^{n} \left ( \left \langle \bm{w}, \phi \left ( \bm{x}_{i}  \right )  \right \rangle  \right ) ^{2} \label{euqation_W_ij}
\end{equation}
The performance of kernel methods critically depends on the choice of the feature map $\phi(\cdot)$, or equivalently, the kernel function $\kappa(\cdot, \cdot)$. A variety of kernels have been proposed for different applications, including radial basis function (RBF) kernels, Gaussian kernels, and quantum kernels.

Compared to classical kernel methods, the key distinction of quantum kernel approaches lies in the use of quantum feature maps that embed classical data into quantum states. For an $N$-qubit system, a data sample $\bm{x}_i$ is encoded via a parameterized quantum circuit $U_c(\bm{x}_i, \bm{\theta})$ acting on the initial state $\lvert 0 \rangle^{\otimes N}$, resulting in the quantum state $\left | \varphi \left ( \bm{x}_{i}, \bm{\theta} \right )    \right \rangle = U_{c}\left ( \bm{x}_{i}, \bm{\theta} \right )  \left | 0  \right \rangle ^{\otimes N}$. This defines an implicit feature map into a high-dimensional Hilbert space. The corresponding density operator is given by $\rho(\bm{x}_i, \bm{\theta}) = \lvert \varphi(\bm{x}_i, \bm{\theta}) \rangle \langle \varphi(\bm{x}_i, \bm{\theta}) \rvert$. A commonly used quantum kernel is defined via the Hilbert–Schmidt inner product, leading to the kernel matrix
\begin{equation}
 \left [ W\left ( \bm{x},\bm{\theta}  \right ) \right ] _{i,j} = Tr\left ( \rho \left ( \bm{x}_{i}, \bm{\theta}   \right ) \rho \left ( \bm{x}_{j}, \bm{\theta}   \right ) \right ) , \forall i,j \in \left [ n \right ] 
\end{equation}
which corresponds to the squared overlap between quantum states for pure-state encodings.
Using this kernel matrix, learning proceeds analogously to classical kernel methods. The performance of a quantum kernel critically depends on the design of the encoding circuit $U_c(\cdot, \bm{\theta})$, which determines the induced feature space. In practice, $U_c$ can take various forms, including predefined ansätze, hardware-efficient circuits, or randomly generated circuits from a given gate set.

\subsection{Graph neural networks}
Graph neural networks have become a powerful framework for learning over graph-structured data, where relationships among entities play a critical role in the underlying data distribution \cite{53}. By iteratively aggregating and transforming information from neighboring nodes through message-passing mechanisms, GNNs are capable of capturing both local topology and global structural dependencies \cite{54}. This capacity to model complex relational patterns has led to their widespread adoption across a variety of domains, including social network analysis \cite{60}, molecular property prediction \cite{he2025self, 61,62,63}, and recommender systems \cite{64,65}. Notably, variants such as Graph Convolutional Networks (GCNs), Graph Attention Networks (GATs), and GraphSAGE have significantly advanced the state-of-the-art in graph representation learning tasks \cite{55,56,57,58,59}. In the context of quantum computing, quantum circuits can be naturally represented as graphs, where qubits correspond to nodes and quantum gates or couplings define the edges. This structural analogy provides a compelling motivation for applying GNNs to quantum systems, enabling efficient learning and prediction tasks grounded in the intrinsic connectivity of quantum architectures.

\section{Method}
\label{Method}

\begin{figure*}[htbp]
	\centering
	\includegraphics[width= 1 \textwidth]{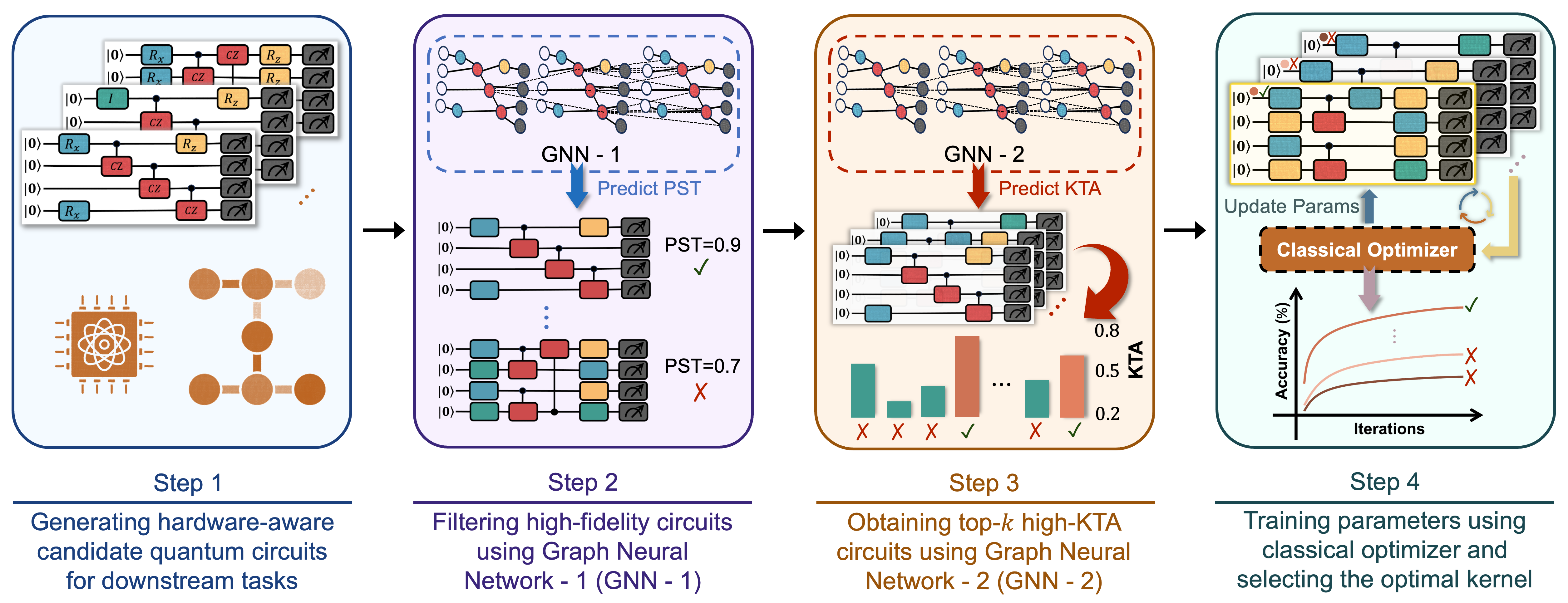}
	\caption{Overview of the proposed framework for hardware-aware quantum kernel design} \label{figure3_a}
\end{figure*}

We now present the proposed framework, illustrated in Fig.~\ref{figure3_a}, which consists of four main steps:

\noindent \textbf{\textit{Step 1 Candidate Circuits Generation:}} A subset of qubits is first selected based on device topology and noise characteristics to ensure favorable connectivity and reduced noise impact. Based on this hardware-aware selection, a large set of candidate quantum circuits is generated under the native gate set and connectivity constraints of the target device.

\noindent \textbf{\textit{Step 2 PST Prediction:}} A graph neural network (GNN-1) is employed to predict PST, which serves as a proxy for circuit fidelity. This enables early rejection of low-fidelity candidates, effectively retaining circuits with higher fidelity potential. 

\noindent \textbf{\textit{Step 3 KTA Prediction:}} A second graph neural network (GNN-2) is designed to estimate the KTA, a surrogate metric for classification accuracy, over the high-fidelity circuits identified in \textbf{\textit{Step 2}}. The top-$k$ circuits with the highest predicted KTA values are selected for further refinement.

\noindent \textbf{\textit{Step 4 Final Training and Kernel Selection:}} A classical optimizer is then employed to train the parameters of the top-$k$ circuits, selected based on high KTA and PST values. The circuit achieving the best task-specific classification performance is chosen as the final quantum kernel.

For clarity, the four-step framework is grouped into three components: candidate circuit generation (\textbf{\textit{Step 1}}, Section~\ref{Candidate_circuit_generation}); GNN-based metric prediction (\textbf{\textit{Steps 2–3}}, Section~\ref{Metric_prediction_based_on_graph_neural_networks}); and final training and kernel selection (\textbf{\textit{Step 4}}, Section~\ref{Final_training_and_kernel_selection}).

\subsection{Candidate circuit generation}
\label{Candidate_circuit_generation}
The generation of candidate quantum circuits consists of three stages. First, a subset of qubits is selected from the quantum device by jointly considering topological connectivity and noise characteristics of both qubits and their couplings, with the goal of identifying a low-noise subgraph that improves hardware efficiency and circuit fidelity. Second, candidate circuits are constructed on the selected qubits under the native gate set and connectivity constraints of the device, where each quantum gate is applied only to physically connected qubits, thereby implicitly satisfying qubit mapping and avoiding the insertion of SWAP gates during compilation, which reduces circuit depth and mitigates noise accumulation. Third, parameterized quantum gates are introduced at different positions within the circuits to enable flexible data embedding, enhancing the expressivity and performance of the resulting quantum kernels.

\subsubsection{Subgraph selection in device topology}
\label{Subgraph_selection_in_device_topology}
\begin{figure*}[htbp]
    \centering
    \includegraphics[width= 1.0 \textwidth]{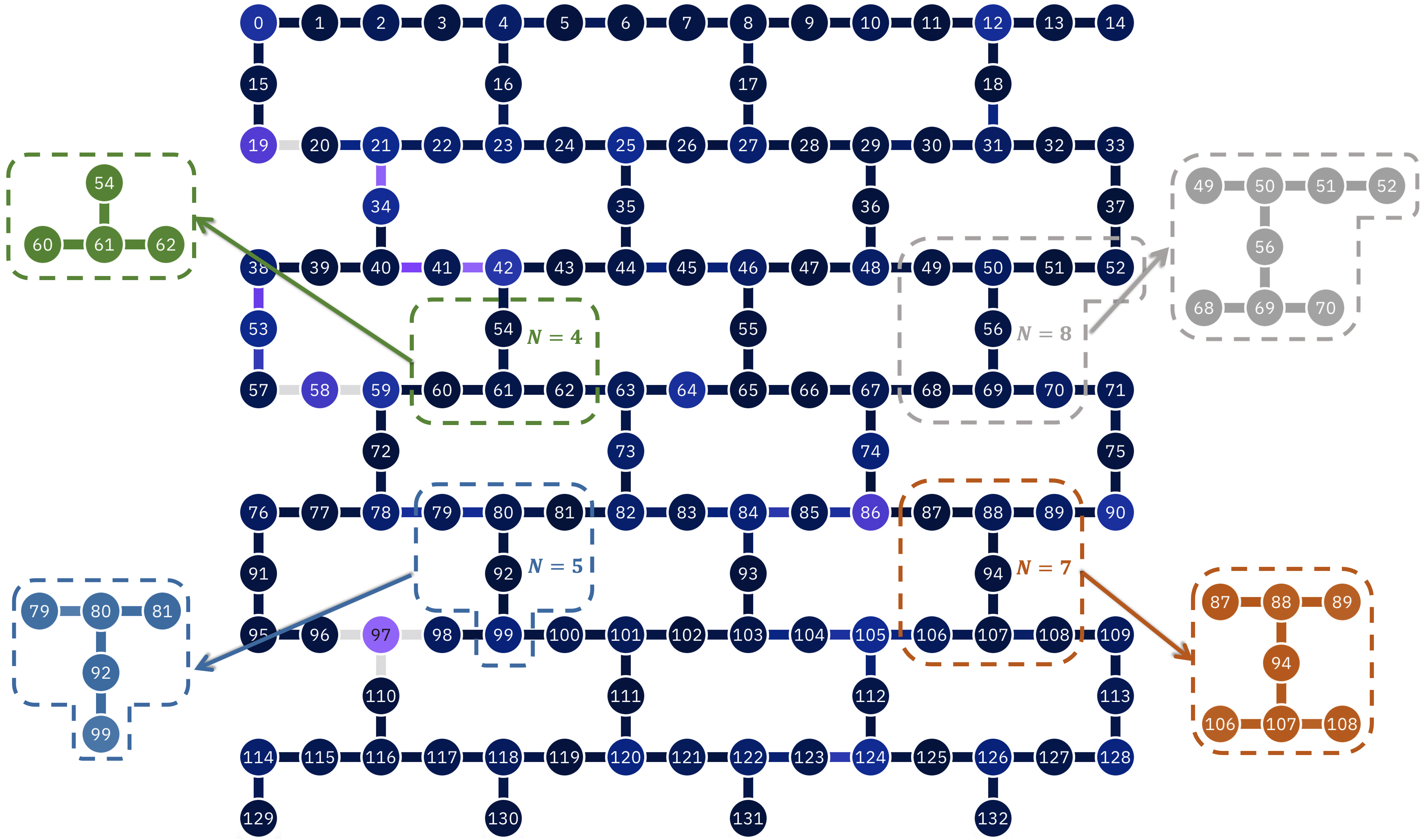}
    \caption{Subgraphs with $N=4,5,7$ and $8$ selected from the 133-qubit IBM Torino topology.} \label{figure3_b}
\end{figure*}

Due to heterogeneous noise levels across qubits and their couplings in quantum devices, we select a subset of $N$ qubits that are less susceptible to noise, thereby improving hardware efficiency and enhancing circuit fidelity. Fig.~\ref{figure3_b} shows example subgraphs with $N=4,5,7,$ and $8$ extracted from the 133-qubit IBM Torino topology.
Selecting a subgraph with $N$ qubits requires jointly considering multiple hardware-specific factors, including single-qubit gate errors, readout errors, two-qubit gate errors between coupled qubits, and qubit connectivity. The objective is to identify a subgraph with high gate fidelity and strong connectivity while maintaining low readout error. By incorporating device-specific noise information, this selection strategy ensures that the resulting quantum circuits are hardware-aware. Detailed procedures are provided in Appendix~\ref{Appendix_A}.

\subsubsection{Generating candidate circuits based on the subgraph}
\label{Generating_candidate_circuits_based_on_the_subgraph}
After selecting a high-fidelity subgraph $G$, we generate $M$ candidate quantum circuits under the structural constraints of $G$ and the native gate set $S$ supported by the device, ensuring hardware-aware circuit construction. As illustrated in \textbf{\textit{Step 1}} of Fig.~\ref{figure3_a}, candidate circuits are generated from a subgraph of the IBM Torino using the gate set $S=\left \{ R_{x}, R_{z}, CZ, I \right \} $. Detailed procedures are provided in Algorithm~\ref{alg1}. During circuit generation, the placement of two-qubit gates between coupled qubits is guided by their gate errors. Specifically, the probability of inserting a two-qubit gate between qubits $q_i$ and $q_j$, denoted as $P(q_i, q_j)$, is defined as follows:
\begin{equation}
\begin{array}{c}
P\left ( q_{i}, q_{j}\mid e_{j}    \right ) = \frac{1/e_{j} }{ {\textstyle \sum_{j} 1/e_{j} } } 
\end{array}
\end{equation}
where the two-qubit gate error $e_{ij}$ corresponds to the edge $q_i q_j$ in the subgraph $G$. Single-qubit gate errors are neglected, as two-qubit gate errors are typically an order of magnitude larger and thus dominate circuit performance. Since circuit construction is explicitly guided by noise information, the resulting circuits are inherently hardware-aware.

After generating the candidate circuits, a lightweight optimization step is applied to fuse consecutive parameterized gates whenever possible, reducing circuit redundancy. By iteratively applying the procedure in Algorithm~\ref{alg1} together with this optimization step, a large pool of hardware-aware candidate circuits can be efficiently generated, providing a strong foundation for subsequent GNN-based evaluation and optimal circuit selection.

\begin{algorithm}
\caption{Generate candidate circuit} \label{alg1}
\begin{algorithmic}[1]
    \Require 
    Device subgraph $G\left ( \left \{ V_{i}   \right \}, \left \{ E_{j} \right \}   \right )$; 2-qubit gate error $\left \{ e_{j}  \right \} $; The gate set $S$; The number of gates $P$. 
    \Ensure
    Candidate circuit $C$.
    \For{$k$ in range($P$)}
    \State Sample a gate $g_k$ uniformly from the set $S$ \Comment{Sample a gate to be added, e.g., $R_{x} $, $R_{z} $, or $CZ$. }
    \If{$g_{k}$ is a 1-qubit gate}
    \State Select a qubit $q_{i}  \in \left \{ V_{i}  \right \} $ uniformly at random from the subgraph $G$ \Comment{Select a qubit from the subgraph as the target qubit of the quantum gate.}
    \State Append $g_{k}\left ( q_{i}  \right )  $ to $C$
    \ElsIf{$g_{k}$ is a 2-qubit gate}
    \State Select a pair of qubits $q_{i},q_{j}  $ from subgraph $G$ according to the probability distribution $P\left ( q_{i}, q_{j}\mid e_{j}    \right ) $  \Comment{2-qubit gate placement is guided by the associated error $e_{j}$.}
    \State Append $g_{k}\left ( q_{i}q_{j}  \right ) $ to $C$
    \EndIf
    \EndFor
    \State        
    \Return $C$
\end{algorithmic}
\end{algorithm}

\subsubsection{Data embedding in circuits}
\begin{figure}[htbp]
	\centering
	\includegraphics[width= 0.6 \textwidth]{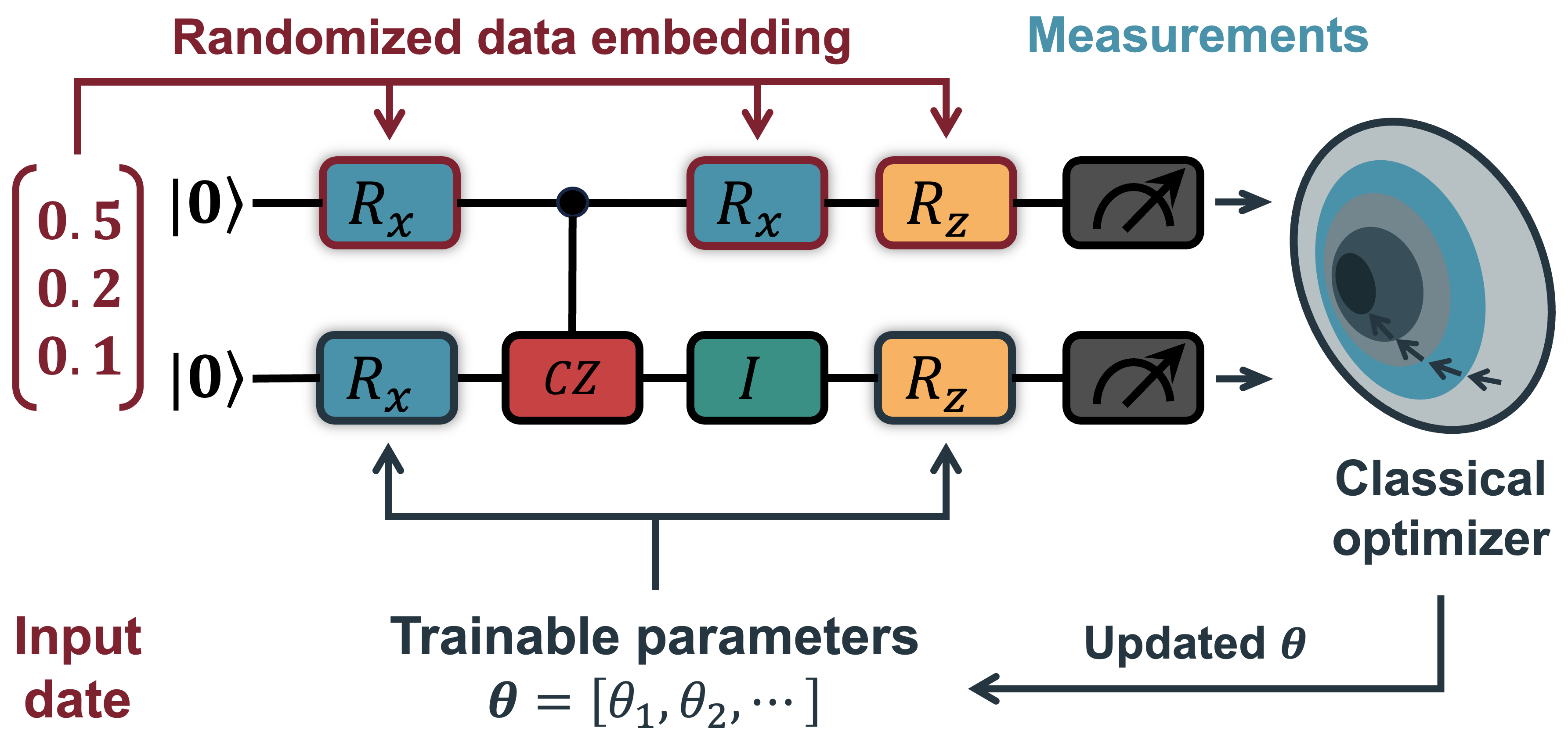}
	\caption{Illustration of randomized data embedding in parameterized quantum circuits.} \label{figure3_b1}
\end{figure}
Inspired by the advantages of randomized data embedding over fixed strategies~\cite{anagolum2024elivagar}, we adopt a simple yet effective embedding scheme, as illustrated in Fig.~\ref{figure3_b1}. Specifically, in each quantum circuit, a subset of parameterized single-qubit gates is randomly selected for data encoding, with each selected gate corresponding to one input feature. To ensure full data coverage, the number of embedding gates is set to be no smaller than the input dimensionality. In addition, we prevent consecutive mergeable parameterized gates from encoding different data, thereby avoiding redundant operations after circuit simplification.

For high-dimensional data, such as image classification tasks (e.g., MNIST with 784 features), direct embedding would require a large number of quantum gates, resulting in deep circuits with many qubits. This not only exceeds the capability of NISQ devices but also exacerbates the vanishing kernel similarity issue, leading to degraded performance. To address this, we incorporate a feature selection step in the preprocessing stage to reduce the input dimensionality. Specifically, the max-relevance and min-redundancy (mRMR) method is employed to project the data from $d$ to $p$ dimensions ($p \ll d$), thereby significantly reducing the number of required embedding gates. This strategy mitigates kernel concentration effects and enables more effective utilization of high-dimensional data on NISQ devices.


\subsection{Metrics prediction based on graph neural networks}
\label{Metric_prediction_based_on_graph_neural_networks}
To predict the probability of successful trials (PST) under device noise and the corresponding performance metric, kernel-target alignment (KTA), we design two graph neural networks (GNNs), referred to as GNN-1 and GNN-2. While the two models share a similar overall architecture, they differ in specific design choices due to their distinct prediction objectives. In the following, we describe the GNN framework in detail, including dataset preparation, graph representation, node features, target definitions, and model architecture.

\subsubsection{Construction of the dataset}
The primary motivation for constructing GNN-based predictors is to enable rapid evaluation of a large number of candidate circuits after training on a relatively small dataset. To this end, we adopt a random sampling strategy to construct the training set, where a subset of circuits is selected from the previously generated pool of candidates. To ensure consistency and completeness, all qubits are initialized in the $\lvert 0 \rangle$ state, and measurement operations are included in each circuit, as illustrated in the left panel of Fig.~\ref{figure3_c}. This formulation also facilitates the incorporation of device-specific noise characteristics, such as readout errors, into the node features.

\subsubsection{Graph construction}
\begin{figure}[htbp]
	\centering
	\includegraphics[width= 0.75 \textwidth]{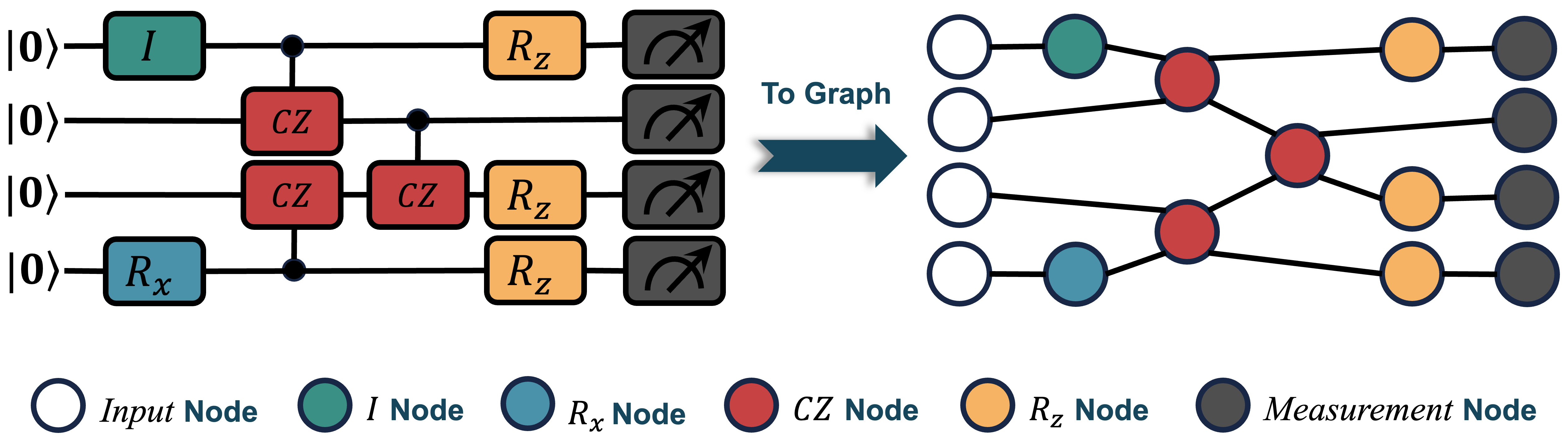}
	\caption{Schematic of quantum circuit-to-graph transformation for GNN-based prediction.} \label{figure3_c}
\end{figure}
Motivated by the structural similarity between quantum circuits and graphs, we represent quantum circuits as directed acyclic graphs (DAGs). In this formulation, nodes correspond to qubits, quantum gates, or measurement operations, while edges encode the temporal ordering of operations from left to right. As illustrated in Fig.~\ref{figure3_c}, circuits generated on IBM Torino are transformed into graph representations, with each two-qubit gate modeled as a single node to preserve its joint operation. This transformation can also be implemented using existing toolkits, such as Qiskit and PyZX, with appropriate configuration.

\subsubsection{Node features}
\begin{figure*}[htbp]
    \centering
    \includegraphics[width= 0.96 \textwidth]{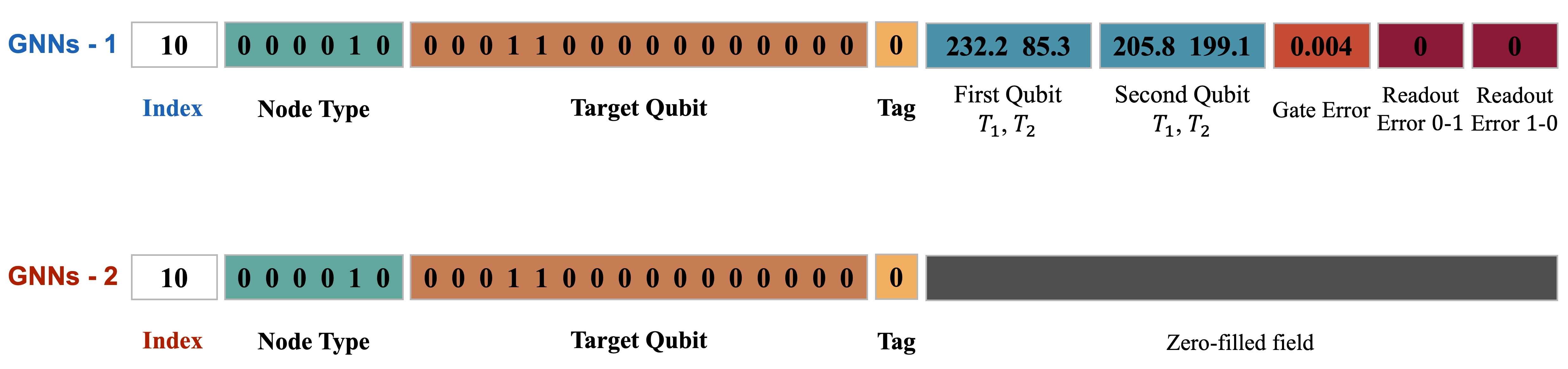}
    \caption{Node feature vectors of GNN-1 and GNN-2 for $CZ$ gate.} \label{figure3_d}
\end{figure*}

For each node in the graph, we construct a feature vector to encode its associated properties. The features include the node index, node type, target qubit, tag, relaxation time ($T_1$), dephasing time ($T_2$) of the target qubit, as well as gate error and readout error. The specific feature configurations used for GNN-1 and GNN-2 are illustrated in Fig.~\ref{figure3_d}.
The feature vectors for both GNN-1 and GNN-2 have a length of 31. It is worth noting that the length of the feature vectors can be manually configured based on either the maximum number of qubits in the circuits or specific requirements, ensuring that each circuit can be properly encoded. In the example shown in Fig. \ref{figure3_d}, we set the maximum supported qubit number to 16, which results in a final feature vector length of 31. The first part of the feature vector encodes the index, which is used for node sorting. The next three parts are one-hot encodings representing the node type, target qubit, and tag, respectively. The node type distinguishes different types of nodes, including initial input qubit, measurement, $R_{x} $ gate, $R_{z} $ gate, $CZ$ gate, and identity ($I$) gate. For example, the one-hot code $000010$ indicates that the node corresponds to a $CZ$ gate. The target qubit labels the index of the qubit affected by the operation, and its length can be expanded according to practical needs. The tag distinguishes whether a gate is used for data embedding, switching from 0 to 1 for embedding gates. In addition to capturing the topological connections between circuits, another advantage of GNNs is their ability to incorporate hardware noise information, which is primarily achieved through node feature vectors. In our method, the last seven elements of each node’s feature vector are dedicated to describing the noise characteristics of the quantum hardware. During model training, GNN-1 incorporates these noise features to predict the fidelity-related metric, allowing the model to implicitly learn how device noise affects circuit performance, and enabling early rejection of poor-performing circuits. In contrast, GNN-2 focuses on the predicting performance-related metric, with the last 7 elements of its feature vectors zero-filled to exclude noise information. The noise parameters are based on real quantum devices such as IBM Torino, IBM Perth, and IBM Nairobi. Specifically, the first 4 elements represent the relaxation time ($T_{1} $) and dephasing time ($T_{2} $) for the two involved qubits, the 5th element encodes the gate operation error, and the final two elements capture the readout errors. For nodes without associated noise characteristics, the corresponding values are set to zero. For instance, measurement nodes have their $T_{1} $, $T_{2} $, and gate error fields set to zero. Fig. \ref{figure3_d} shows the complete feature vector representation of a $CZ$ gate node for both GNN-1 and GNN-2.

\subsubsection{Target values}
\begin{figure}[htbp]
    \centering
    \includegraphics[width= 0.75 \textwidth]{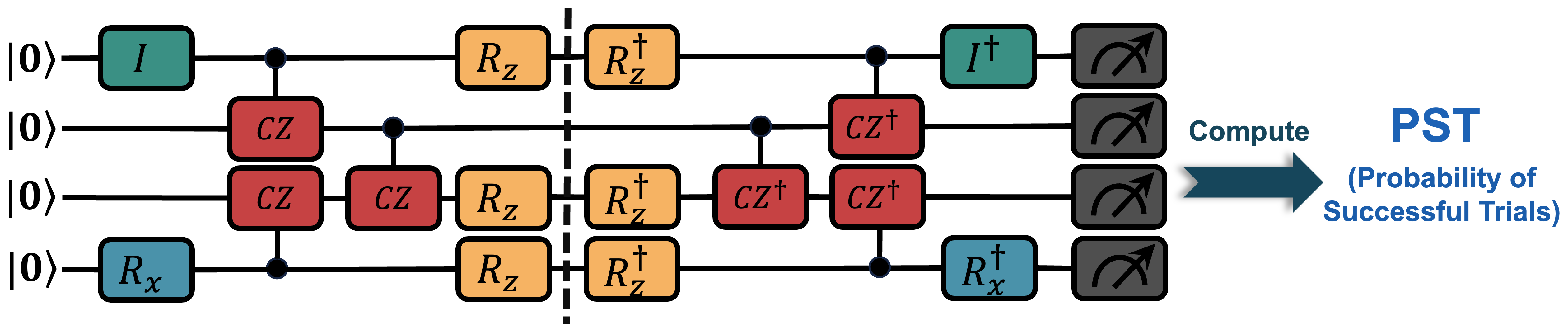}
    \caption{Schematic illustration of PST estimation via circuit inversion.} \label{figure3_e}
\end{figure}

To evaluate the impact of noise on quantum circuits, we consider fidelity, a standard metric that quantifies the closeness between quantum states. In the presence of noise, fidelity measures the deviation between the state produced by a noisy circuit and its noiseless counterpart. A large deviation indicates that the circuit is significantly affected by noise and may yield unreliable outputs. However, estimating fidelity typically requires quantum state tomography, which is computationally expensive. To circumvent this limitation, we adopt the probability of successful trials (PST)~\cite{wang2022graph} as a proxy metric, which has been shown to provide accurate estimates of circuit fidelity. The PST is defined as follows:
\begin{equation}
PST=\frac{T_{initial}}{T_{total} } 
\end{equation}
where $T_{\text{initial}}$ denotes the number of trials that return the initial state $\lvert 0 \rangle^{\otimes N}$, and $T_{\text{total}}$ is the total number of trials. Rather than directly estimating fidelity, we append the inverse of each circuit to itself and apply the resulting circuit to the initial state $\lvert 0 \rangle^{\otimes N}$. The PST is then obtained as the fraction of measurement outcomes corresponding to the initial state, as illustrated in Fig.~\ref{figure3_e}. The resulting PST values are used as target labels for training GNN-1.

GNN-2 is designed to predict the kernel-target alignment (KTA), a metric that has been shown to correlate positively with training accuracy~\cite{lei2024neural}, and thus serves as a reliable surrogate for model performance. Before introducing KTA, we briefly review the concept of kernel alignment, originally proposed in~\cite{cristianini2001kernel}, which measures the similarity between two kernel matrices $K_1$ and $K_2$:
\begin{equation}
KA\left ( K_{1}, K_{2}   \right )=\frac{\left \langle {K_{1}, K_{2}}   \right \rangle_{F} }{\sqrt{\left \langle {K_{1}, K_{1}}   \right \rangle_{F} \left \langle {K_{2}, K_{2}}   \right \rangle_{F} } }
\end{equation}
where $\left \langle K_{1},K_{2}  \right \rangle _{F} $ denotes the Frobenius inner product between two matrices, which is given by
\begin{equation}
\left \langle K_{1},K_{2}  \right \rangle _{F} =\sum_{i=1}^{l} \sum_{j=1}^{l} \kappa_{1}\left (x_i,x_j  \right ) \kappa_{2}\left (x_i,x_j  \right )
\end{equation}
where $l$ denotes the size of the kernel matrix, and $\kappa(x_i, x_j) = [K]_{i,j}$ represents the $(i,j)$-th entry of the kernel matrix $K$. For classification tasks, an ideal target matrix is defined as $K^{\ast} = yy^{T}$, where $y = (y_1, y_2, \cdots, y_l)^{T}$ denotes the label vector of the training set $\mathcal{D}$. The kernel-target alignment (KTA) is then defined as the alignment between the kernel matrix $K$ (e.g., induced by the quantum kernel) and the target matrix $K^{\ast}$:
\begin{equation}
KTA\left ( K \right ) = KA\left ( K, K^{\ast }  \right ) = \frac{\left \langle K,yy^{T}  \right \rangle _{F}  }{ \sqrt{ \left \langle K,K  \right \rangle _{F} \left \langle yy^{T},yy^{T}  \right \rangle _{F} }}  \label{euqation_KTA}
\end{equation}
Since $\left \langle A,B \right \rangle_{F}= Tr\left ( A^{T}B  \right ) $ and $K^{T}=K$, substituting them into Eq. \ref{euqation_KTA} yields the following formula:
\begin{equation}
KTA\left ( K \right ) =\frac{Tr\left ( Kyy^{T}  \right ) }{\sqrt{Tr\left ( K^{2}  \right ) Tr\left ( \left ( yy^{T}  \right ) ^{2}  \right )} } =\frac{y^{T}Ky }{l\sqrt{Tr\left ( K^{2}  \right ) } } 
\end{equation}
Here, $l$ denotes the length of the label vector $y$. For multi-class classification tasks (including the binary case), the ideal target matrix $K^{\ast}$ can be constructed as follows~\cite{camargo2009multi}:
\begin{equation}
\left [ K^{\ast }  \right ] _{i,j} = \begin{cases}
 1 & y_{i} = y_{j}   \\
\frac{-1}{c-1} & y_{i} \ne  y_{j}
\end{cases}
\end{equation}
where $c$ ($c \geq 2$) denotes the number of classes. As KTA serves as a reliable surrogate for classification accuracy and can be efficiently computed given a kernel matrix, we adopt it as the performance metric for quantum circuits. However, constructing kernel matrices from quantum circuits remains computationally expensive, motivating the use of GNN-2 to directly predict KTA.

\begin{figure*}[htbp]
	\centering
	\includegraphics[width= 0.98 \textwidth]{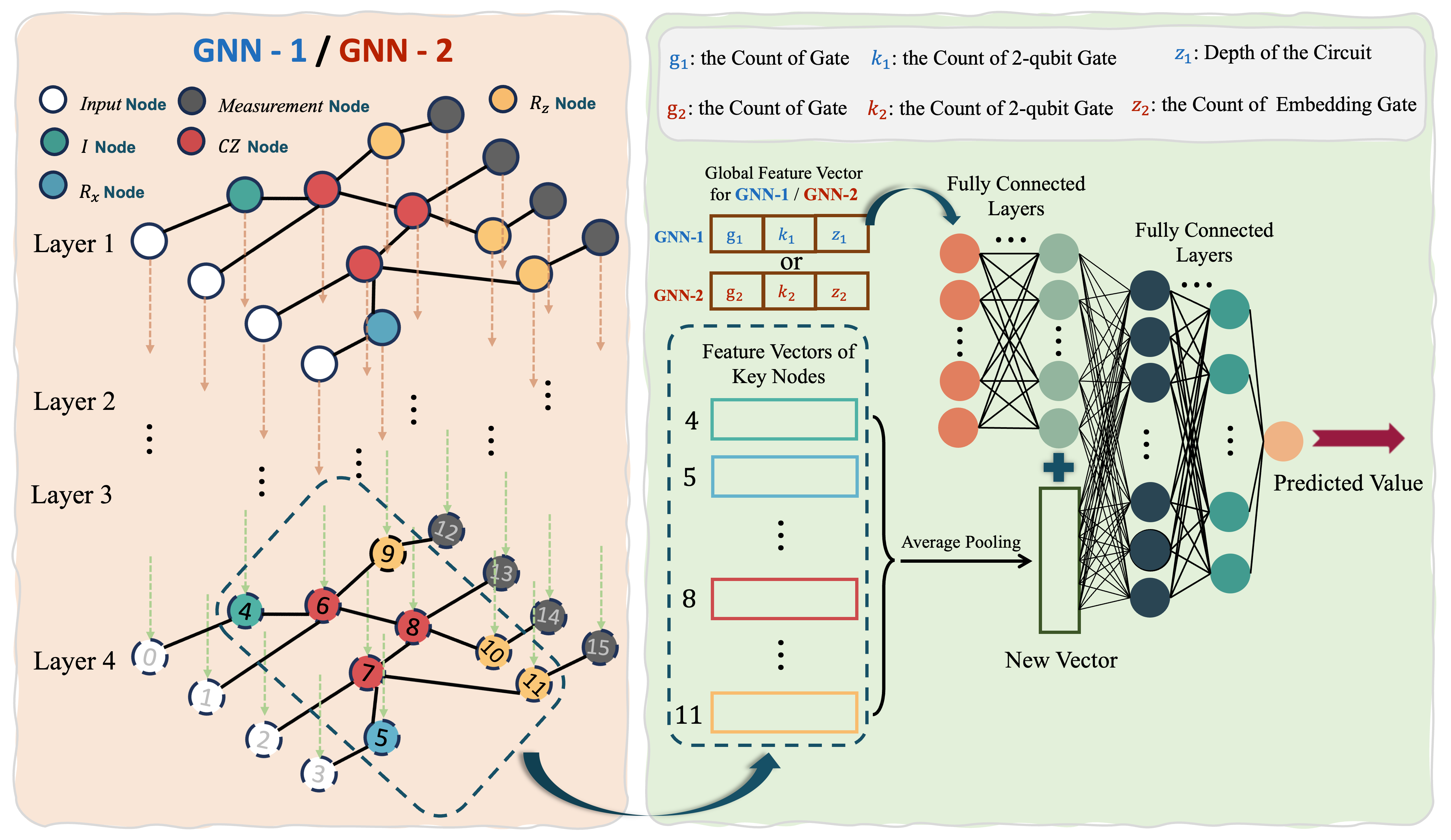}
	\caption{Architecture of GNN-based predictors for quantum circuit evaluation.} \label{figure3_f}
\end{figure*}
\subsubsection{Construction of GNNs}
We adopt similar graph neural network (GNN) architectures for GNN-1 and GNN-2, as illustrated in Fig.~\ref{figure3_f}. Both models employ a four-layer message-passing structure, with each layer augmented by an attention mechanism to adaptively weight edges during propagation, thereby enhancing the representation of circuit characteristics. LeakyReLU is used as the activation function between layers.

After graph-based feature extraction, average pooling is applied over the representations of key nodes, excluding input and measurement nodes. The pooled features are then concatenated with the output of a three-layer fully connected network that processes global circuit features. The resulting representation is fed into a multilayer perceptron (MLP) to produce the final prediction.

While the overall architectures are shared, GNN-1 and GNN-2 differ in their global feature inputs. Specifically, GNN-1 uses $(g_1, k_1, z_1)$, where $g_1$ denotes the total number of gates, $k_1$ the number of two-qubit gates, and $z_1$ the circuit depth. In contrast, GNN-2 uses $(g_2, k_2, z_2)$, where $g_2$ denotes the total number of gates, $k_2$ the number of two-qubit gates, and $z_2$ the number of embedding gates. These task-specific global features align with the distinct prediction objectives of the two models, thereby improving predictive performance.

\subsection{Final training and kernel selection}
\label{Final_training_and_kernel_selection}
This step aims to identify the most effective quantum kernel for the target classification task. Based on the results of \textbf{\textit{Steps 2--3}}, we select the top-$k$ candidate circuits with high predicted PST and KTA values. These circuits are then trained on the target task, where the circuit structure is fixed and only the parameters are optimized from scratch using the Adam optimizer.
After training, the classification performance of each circuit is evaluated on the test set, and the circuit achieving the highest accuracy is selected as the final quantum kernel.

\section{Experiment}
\label{Evaluation}
In this section, we present a comprehensive experimental evaluation of the proposed framework. The study is organized into two parts. The first evaluates the predictive performance and computational efficiency of the proposed graph neural networks, GNN-1 and GNN-2 (Section \ref{Predictive_performance_of_the_graph_neural_networks}). The second assesses the overall performance of our methodology against recent state-of-the-art methods across multiple datasets (Section \ref{Performance_of_HaQGNN_on_benchmarks}).

\subsection{Predictive performance of the graph neural networks}
\label{Predictive_performance_of_the_graph_neural_networks}
\noindent \textbf{Model and Training Setups.} In the default setup, both GNN-1 and GNN-2 adopt a four-layer graph neural network architecture with attention mechanisms integrated into each layer, following the design of Graph Attention Networks (GATs). This enables adaptive weighting of neighboring nodes during message passing. At the final layer, average pooling is applied over key nodes (excluding input and measurement nodes) to obtain a 31-dimensional representation. This representation is concatenated with a global feature vector produced by a three-layer fully connected network, where each hidden layer has 12 units.

The concatenated vector is then passed through a multilayer perceptron (MLP) with hidden layer sizes of 256, 128, and 64 to generate the final prediction. LeakyReLU with $\alpha = 0.02$ is used as the activation function throughout the network. Training is performed using the Adam optimizer with a learning rate of 0.01, a batch size of 512, and the mean squared error (MSE) as the loss function for 200 epochs. During preprocessing, node features are normalized using Min–Max scaling across the dataset. The predictive performance is evaluated using the coefficient of determination ($R^2$), defined as:
\begin{equation}
R^{2}=1-\frac{ {\textstyle \sum_{i=1}^{N } }  \left ( y_{i} - \hat{y} _{i}\right ) ^{2}  }{ {\textstyle \sum_{k=1}^{N } } \left ( y_{i} - \bar{y}  \right ) ^{2} } 
\end{equation}
where $\hat{y}_i$ denotes the predicted value for the $i$-th sample, $y_i$ is the corresponding ground-truth value, $\bar{y}$ is the mean of the ground-truth values, and $N$ is the number of samples in the test set.

\noindent \textbf{Dataset Setup.} Following the methodology described in Section~\ref{Candidate_circuit_generation}, we generate 50,000 candidate circuits, from which 5,000 are randomly sampled to construct the dataset for training GNN-1 and GNN-2. Among these, 4,000 circuits are used for training and the remaining 1,000 for testing. Although surrogate metrics (PST and KTA) are employed to reduce the cost of label generation, their large-scale computation remains time-consuming, a trade-off that will be further examined in the subsequent runtime analysis. 

For GNN-1, the noise models used during training are derived from real quantum hardware, including several 7-qubit devices (IBM Perth, IBM Lagos, IBM Nairobi, and IBM Jakarta) as well as the 133-qubit IBM Torino. The considered noise sources include gate errors, relaxation time ($T_1$), dephasing time ($T_2$), and readout errors. In addition, all 7-qubit devices share the same native gate set $S^{\ast} = { R_z, X, \text{CNOT}, I }$.

To evaluate the predictive performance of GNN-1 and GNN-2 for PST and KTA prediction, respectively, we conduct simulations across different system sizes with $N = 4, 5, 7,$ and $8$ qubits. Although these scales are relatively small, they are sufficient to demonstrate the effectiveness of the proposed approach under NISQ-relevant conditions. Extending the framework to larger quantum systems will be the subject of future work, potentially leveraging tensor-network-based simulations~\cite{33} and direct evaluations on real quantum hardware.

\begin{figure}[htbp]
  \centering
  \begin{subfigure}{0.5\textwidth}
    \centering
    \includegraphics[width=\linewidth]{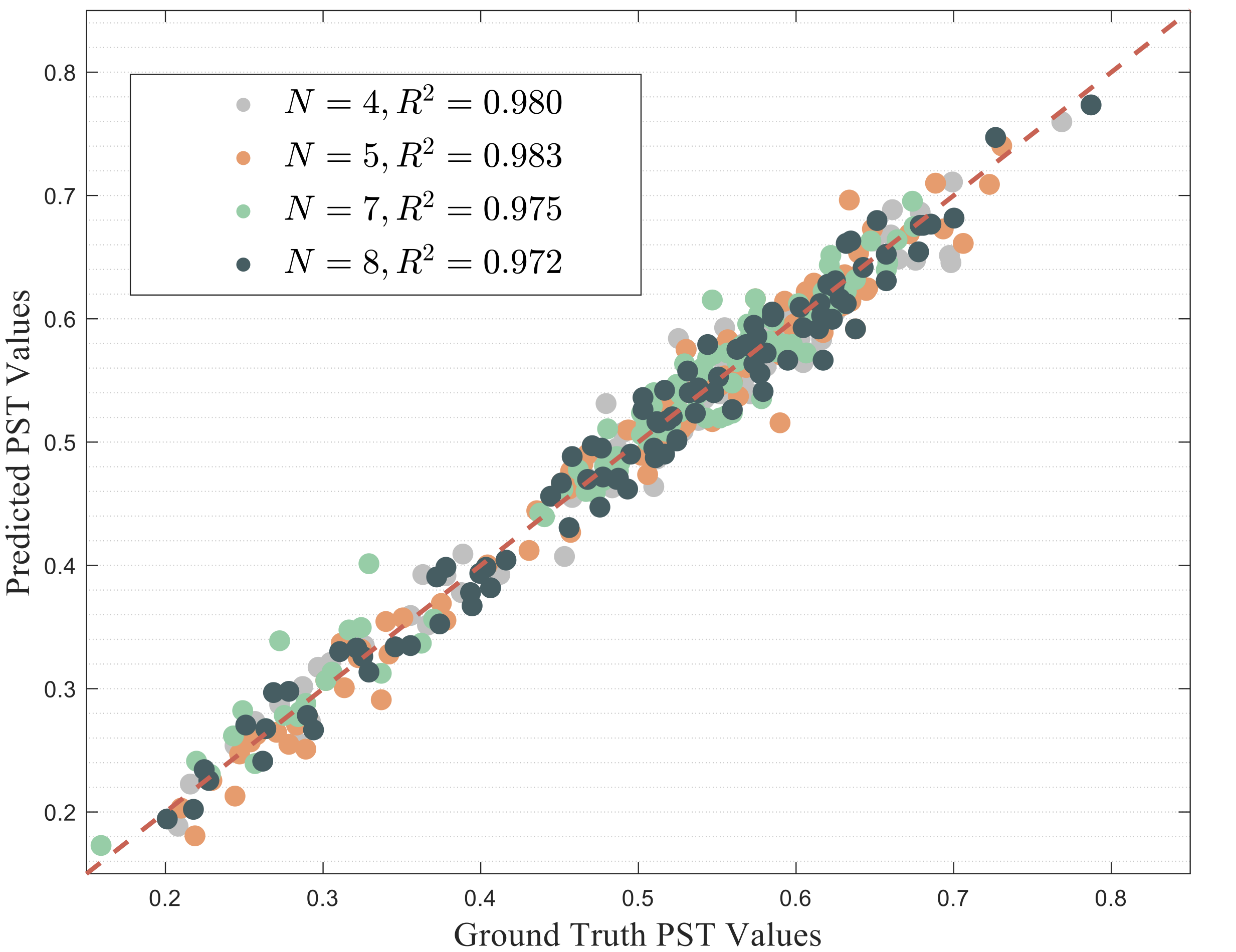}
    \caption{Scatter plot of predicted PST against ground truth values}
    \label{fig:4a}
  \end{subfigure}%
  \hfill
  \begin{subfigure}{0.5\textwidth}
    \centering
    \includegraphics[width=\linewidth]{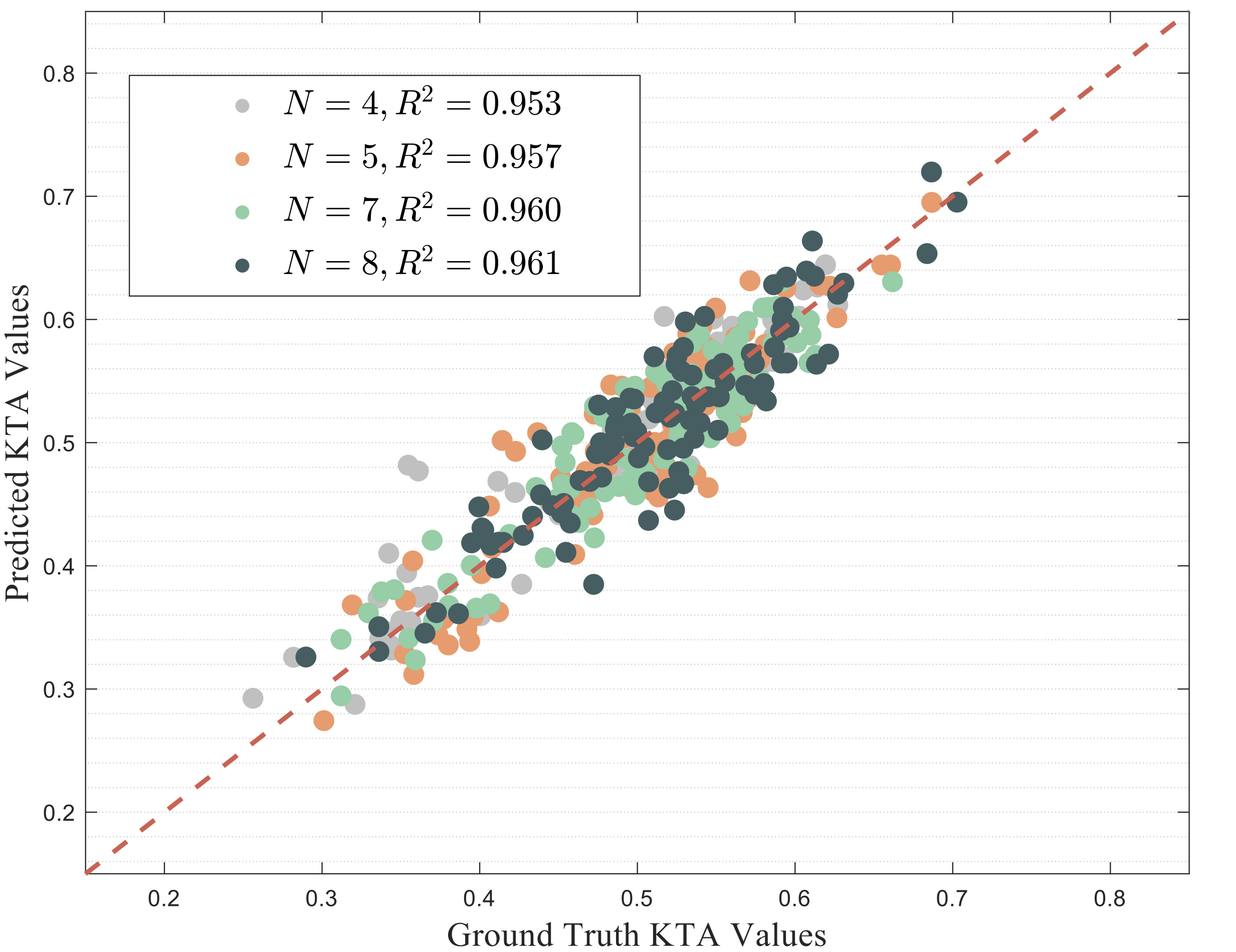}
    \caption{Scatter plot of predicted KTA against ground truth values}
    \label{fig:4b}
  \end{subfigure}
  
  \caption{Scatter plots of predicted PST and KTA values with $N=4,5,7$ and $8$.}
  \label{figure4_ab}
\end{figure}

As shown in Fig.~\ref{figure4_ab}, simulations are carried out on corresponding quantum devices: IBM Perth for $N=4$, and IBM Torino for $N=5, 7,$ and $8$. For visualization purposes, 100 samples are randomly selected from the test set in each case, while the coefficient of determination ($R^2$) is computed over the full test set.

The results in Fig. \ref{figure4_ab} demonstrate that both GNN-1 and GNN-2 achieve high prediction accuracy, with $R^{2} $ exceeding 0.97 for PST and 0.95 for KTA. The high accuracy of GNN-1 in predicting PST under hardware noise is attributed to the inclusion of noise information in the node feature vectors, which enhances the model’s ability to learn noise-informed representations. Moreover, we observe that KTA values are tightly clustered, while PST values exhibit a broader distribution. This implies that quantum kernels with similar KTA values, i.e., theoretically similar classification performance, can exhibit significantly different results on real hardware due to noise. This observation further justifies our strategy of early rejection of low-fidelity candidate circuits based on PST prediction.

\begin{figure}[htbp]
	\centering
	\includegraphics[width= 0.75 \textwidth]{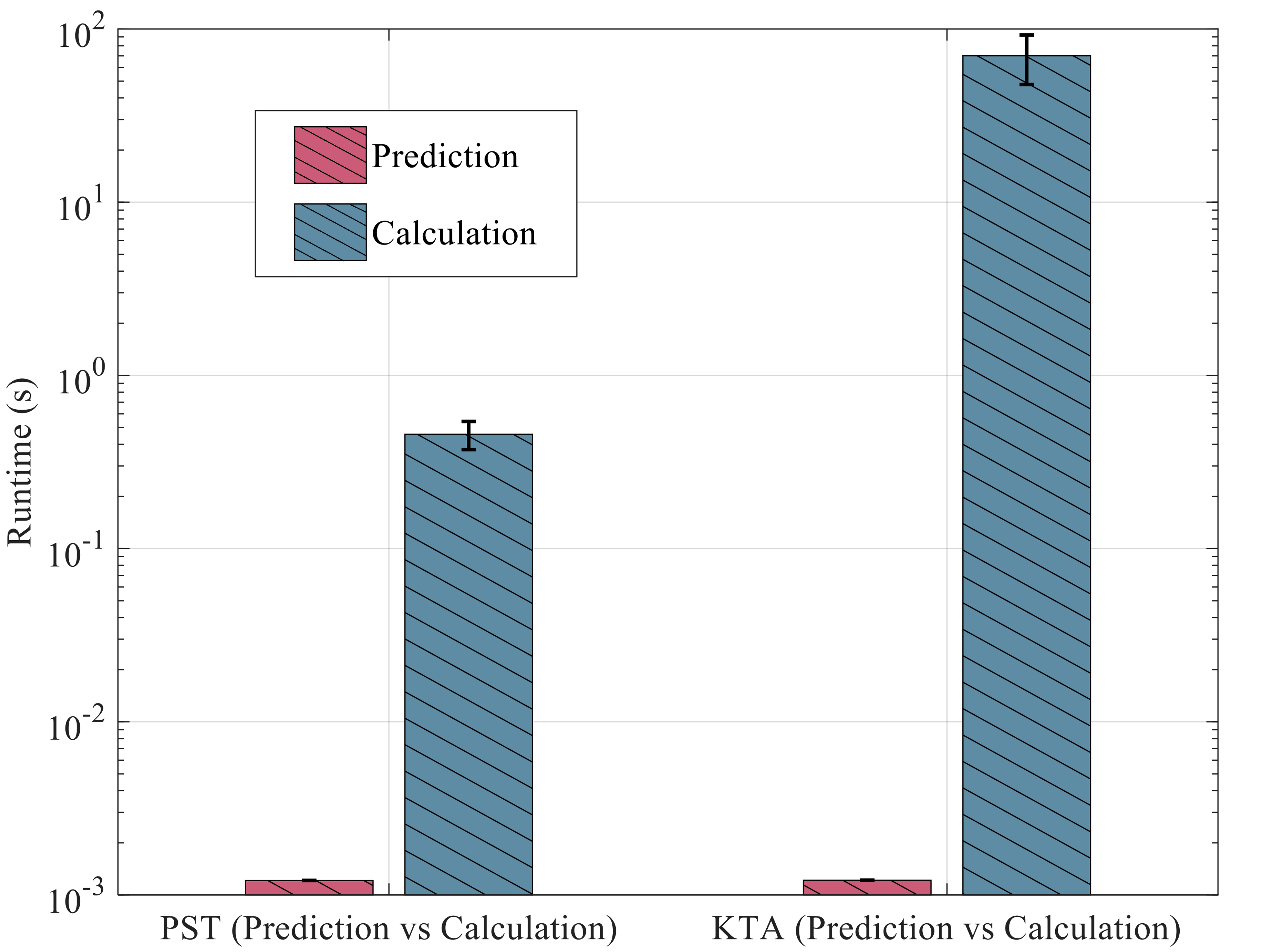}
	\caption{Comparison of runtime of prediction and direct calculation for PST and KTA.} \label{figure4_c}
\end{figure}

To further demonstrate the efficiency of our approach, we evaluate the computational speedup achieved by GNN-1 and GNN-2 in circuit performance estimation. Specifically, we compare the inference time for 100 GNN-1 predictions of PST with the time required for direct PST computation, and similarly compare the inference time for 100 GNN-2 predictions of KTA with that of direct KTA evaluation.
As shown in Fig.~\ref{figure4_c} and summarized in Table~\ref{table1}, the results clearly highlight the substantial efficiency gains of the proposed approach. Note that the y-axis in Fig.~\ref{figure4_c} is presented on a logarithmic scale.
\begin{table}[!htbp]
    \centering
    \caption{Runtime of Prediction and Direct Calculation for PST and KTA} \label{table1}
    \begin{tabular}{c|ccc} 
    \toprule
    Metric   &  Prediction ($\cdot s$)   & Calculation ($\cdot s$)   &   Speedup \\
    \midrule
    \midrule
    PST      &   $0.0012\pm 0.0000$      &   $0.4572\pm 0.0845$      &   $\bm{381\times }$    \\ 
    \midrule
    KTA      &   $0.0012\pm 0.0000$      &   $70.0540\pm 22.2887$    &   $\bm{58378\times }$    \\ 
    \bottomrule
    \end{tabular}
 \end{table}
Compared to direct computation, GNN-1 and GNN-2 achieve speedups of $381\times$ and $58{,}378\times$ for PST and KTA prediction, respectively. This comparison reflects the inference cost of the trained GNNs relative to the expense of explicit computation. Rather than a direct algorithmic acceleration, the improvement arises from replacing costly evaluations with learned surrogate models, enabling rapid performance estimation once training is completed.

Such a paradigm is particularly advantageous when direct computation becomes prohibitively expensive, allowing efficient evaluation of a large number of candidate circuits and making large-scale quantum kernel screening feasible. In addition to computational efficiency, we further examine the scalability of GNN-1 and GNN-2 with respect to the number of qubits, with detailed results provided in Appendix~\ref{Appendix_B}.

\subsection{Performance of Our Framework on Benchmarks}
\label{Performance_of_HaQGNN_on_benchmarks}
\noindent \textbf{Model and Training Setups.} After evaluating candidate circuits using GNN-1 and GNN-2, we select the top-$k$ circuits with high KTA and PST values. In our experiments, $k=10$, corresponding to the top 10 high-performing quantum kernels. These circuits are then fine-tuned by optimizing their parameters from scratch. The training objective is to maximize classification accuracy using support vector machines (SVMs) constructed from the corresponding quantum kernels. Parameter optimization is performed using the Adam optimizer with a learning rate of 0.01.

We further evaluate the proposed method across multiple noisy simulators incorporating hardware-calibrated noise models derived from IBM Perth, IBM Lagos, IBM Nairobi, IBM Jakarta, and IBM Torino. Performance is compared against seven representative baselines. Each method is evaluated over 10 independent runs, and the results are reported as mean values with error bars indicating the standard deviation (SD):

\textbf{Random Quantum Kernel:} A total of 25 quantum kernels are randomly sampled from the gate set, with their circuit depths matched to those generated by the proposed method. Each kernel is trained independently, and the one achieving the highest classification accuracy is selected as the representative quantum kernel for each simulation run.

\textbf{Training Embedding Kernels (TEK) \cite{100}:} This baseline adopts a fixed circuit structure with parameter optimization. In our setup, each TEK kernel is composed of six blocks, including encoding blocks and training blocks. Each block consists of two layers of single-qubit gates followed by a layer of two-qubit gates arranged in a ring topology. Specifically, on 7-qubit devices, the single-qubit layers comprise an $X$ layer (applying $X$ gates to all qubits) followed by an $R_z$ layer, whereas on the IBM Torino device, an $R_x$ layer is used instead of the $X$ layer, followed by an $R_z$ layer.

In the encoding blocks, dimension-reduced and normalized input features are embedded into the circuit by assigning them to the phases of parameterized gates after scaling by $\pi$. Once all features have been encoded, any remaining parameterized gates continue the embedding process by reusing the feature vector in a cyclic manner. The parameters in the training blocks are subsequently optimized to further refine the behavior and performance of the resulting quantum kernel.

\textbf{Radial Basis Function Kernel (RBFK) \cite{101,102}:} A widely used classical kernel with the kernel matrix defined as $\left [ K  \right ]_{i,j} = exp\left ( -\gamma \left \| x_{i}-x_{j}   \right \| ^{2}  \right )  $, where $\gamma $ is a hyperparameter. In the simulations, $\gamma $ is set to the reciprocal of the input feature vector dimension, which serves as a reasonable heuristic to balance the sensitivity of the kernel.

\textbf{QuantumSupernet \cite{du2022quantum}:} Most hyperparameters follow the settings in \cite{du2022quantum}, with one modification: we replace full-batch gradient descent with mini-batch gradient descent (batch size = 32) to ensure efficient and thorough parameter updates during SuperCircuit training. During the construction of the SuperCircuit, we prohibit the occurrence of consecutive quantum gates that can be merged or canceled, such as repeated 1-qubit gates of the same type applied consecutively.

\textbf{QuantumNAS \cite{wang2022quantumnas}:} We adopt the same hyperparameter settings as in \cite{wang2022quantumnas} for both SuperCircuit training and evolutionary search. The gate set is adapted to match the constraints of each target quantum device.

\textbf{EliVagar \cite{anagolum2024elivagar}:} This method also incorporates early rejection of low-fidelity circuits. We follow the hyperparameter settings provided in \cite{anagolum2024elivagar} and adapt the gate set to match the supported operations of the target quantum devices.

\textbf{QuKerNet \cite{lei2024neural}:} This baseline is a recent neural predictor-based approach for quantum kernel design. In our experiments, we follow the hyperparameter settings reported in \cite{lei2024neural} and simulate the circuits using noise models derived from real quantum hardware. The method employs a multilayer perceptron (MLP) to predict the kernel-target alignment (KTA) of quantum circuits represented in matrix form.

Data encoding follows a sequential scheme, where an initial subset of parameterized gates is used to embed the input features, while the remaining parameters are optimized during training. Similar to our approach, a top-$k$ selection strategy is adopted, with the same value of $k$ used for fair comparison.

\begin{figure*}[htbp]
	\centering
	\includegraphics[width= 0.88 \textwidth]{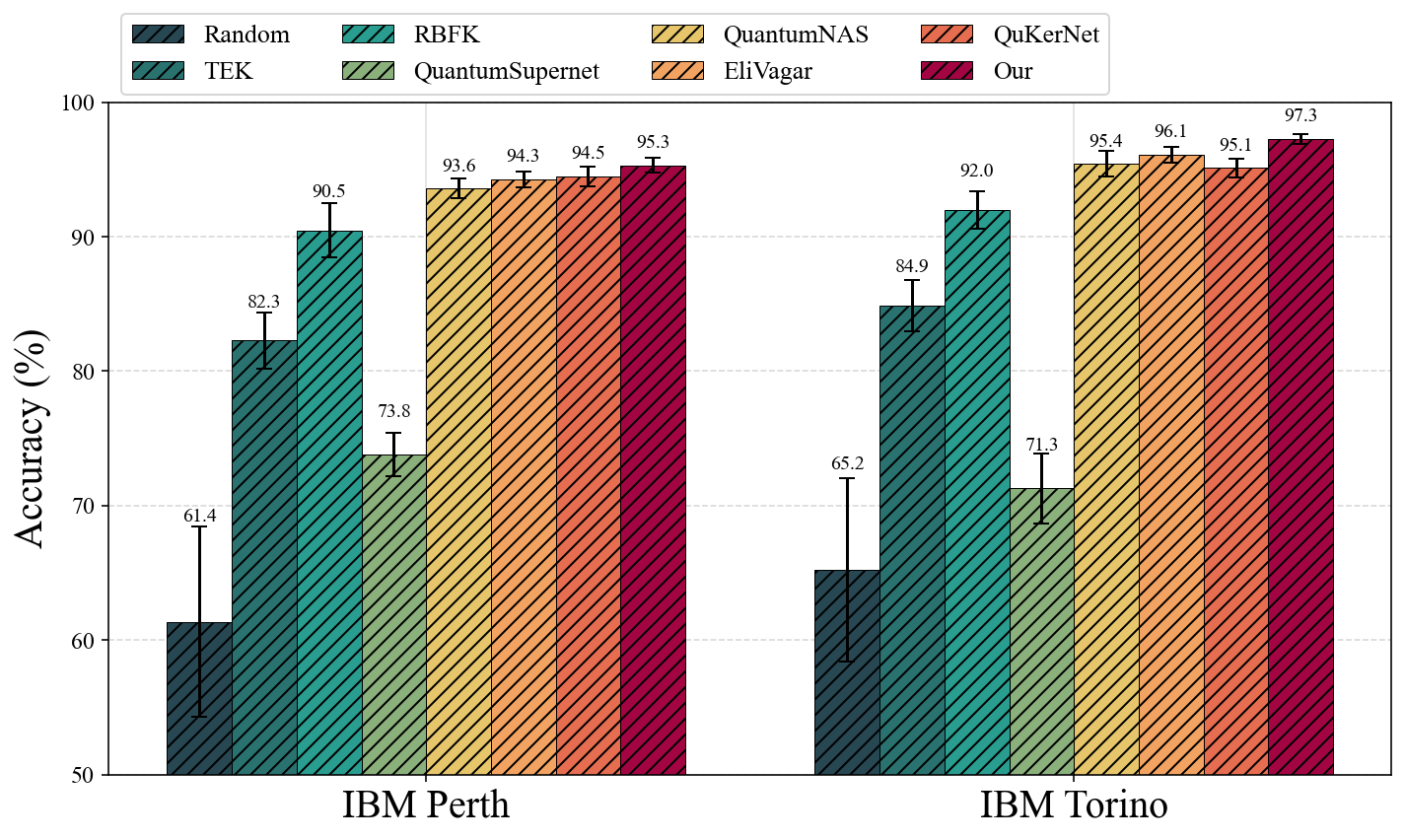}
	\caption{Classification accuracy of different methods on the Credit Card (CC) task under noisy simulation.} \label{figure4_d}
\end{figure*}
In all simulation experiments, baseline methods are configured to use the same number of qubits and the same gate set as the proposed method to ensure fair comparison. When the logical circuits generated by these baselines are incompatible with the target hardware, Qiskit’s default mapping and routing strategies are applied for compilation. Specifically, logical qubits are mapped to physical qubits in a sequential manner, and SWAP gates are heuristically inserted to satisfy hardware connectivity constraints.

\noindent \textbf{Dataset Setup.} We evaluate the proposed method on three quantum machine learning (QML) benchmark tasks: Credit Card (CC) fraud detection (binary classification), MNIST-5 (five-class classification over digits 0–4), and FMNIST-4 (four-class classification over T-shirt, Trouser, Bag, and Ankle Boot).

For MNIST-5 and FMNIST-4, balanced datasets are constructed by selecting 2,000 training samples and 400 test samples from each dataset. The original 784-dimensional images are reduced to 40 dimensions using the minimum Redundancy Maximum Relevance (mRMR) algorithm~\cite{lei2024neural}.

For the CC task, 1,000 samples are drawn while ensuring that at least 5\% correspond to fraudulent transactions. The dataset is randomly split into training and test sets with a 70/30 ratio. The original 28-dimensional features are reduced to 14 dimensions using mRMR.

To validate the performance advantage of the quantum kernels discovered by our framework, we conduct experiments on multiple benchmark tasks under different hardware configurations.

For the 4-qubit Credit Card (CC) task, experiments are performed on two IBM quantum devices with different scales: a 7-qubit device (IBM Perth) and a 133-qubit device (IBM Torino), where only a 4-qubit subset is used in each case. As shown in Fig.~\ref{figure4_d}, the proposed methodology achieves the highest classification accuracy among all compared methods, with more pronounced improvements on the larger IBM Torino device. This gain stems from the hardware-aware design, which enables the selection of low-noise qubits and the generation of high-fidelity circuits tailored to device-specific characteristics, particularly beneficial on larger-scale quantum processors.

\begin{figure*}[htbp]
    \centering
    \includegraphics[width= 1.1 \textwidth]{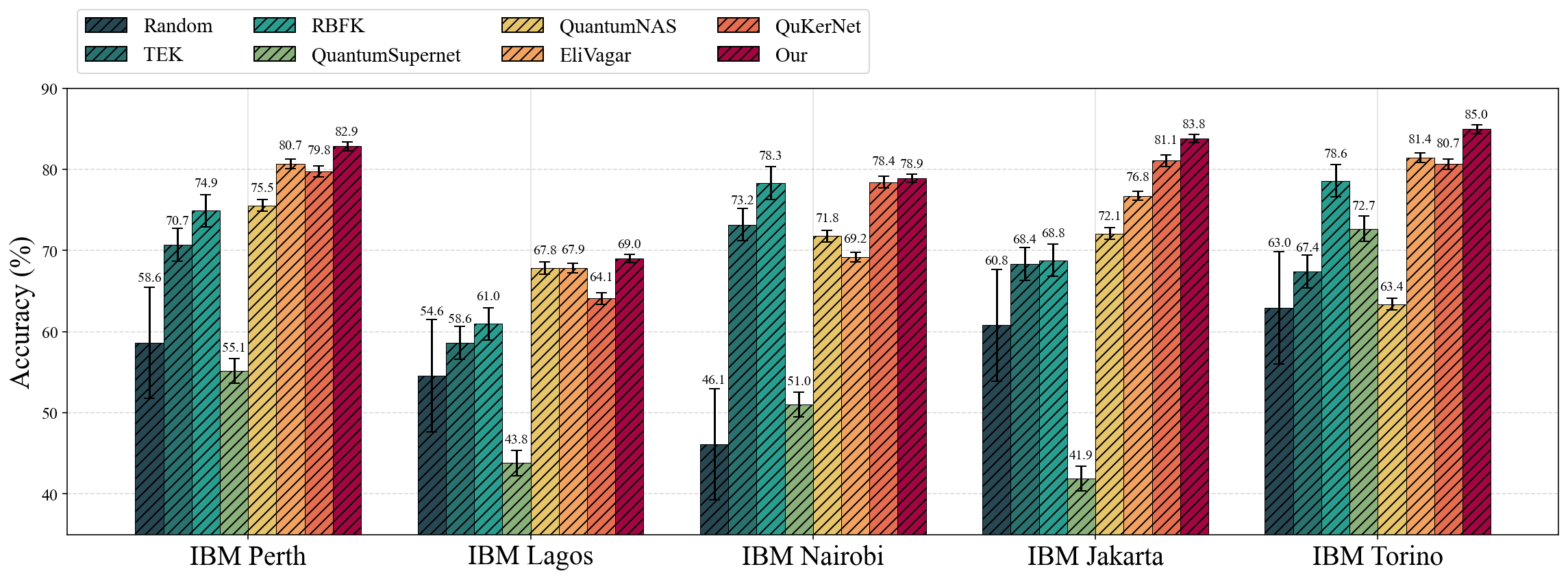}
    \caption{Classification accuracy of different methods on the MNIST-5 classification task under noisy simulation.} \label{figure4_e}
\end{figure*}

We further evaluate the 7-qubit MNIST-5 task on five IBM quantum backends: IBM Perth, IBM Lagos, IBM Nairobi, IBM Jakarta, and IBM Torino (using a 7-qubit subset). As shown in Fig.~\ref{figure4_e}, our method consistently outperforms all baselines in terms of classification accuracy, while also exhibiting improved stability as indicated by smaller error bars. On 7-qubit devices, the limited hardware scale constrains the effectiveness of hardware-aware qubit selection due to the prevalence of noisy qubits. Nevertheless, the proposed framework maintains strong performance owing to its accurate circuit evaluation via GNN-based predictors. On the larger IBM Torino device, the performance advantage becomes more significant, as the method can better exploit hardware heterogeneity to identify low-noise qubits and construct higher-quality circuits.

\begin{figure}[htbp]
	\centering
	\includegraphics[width= 0.75 \textwidth]{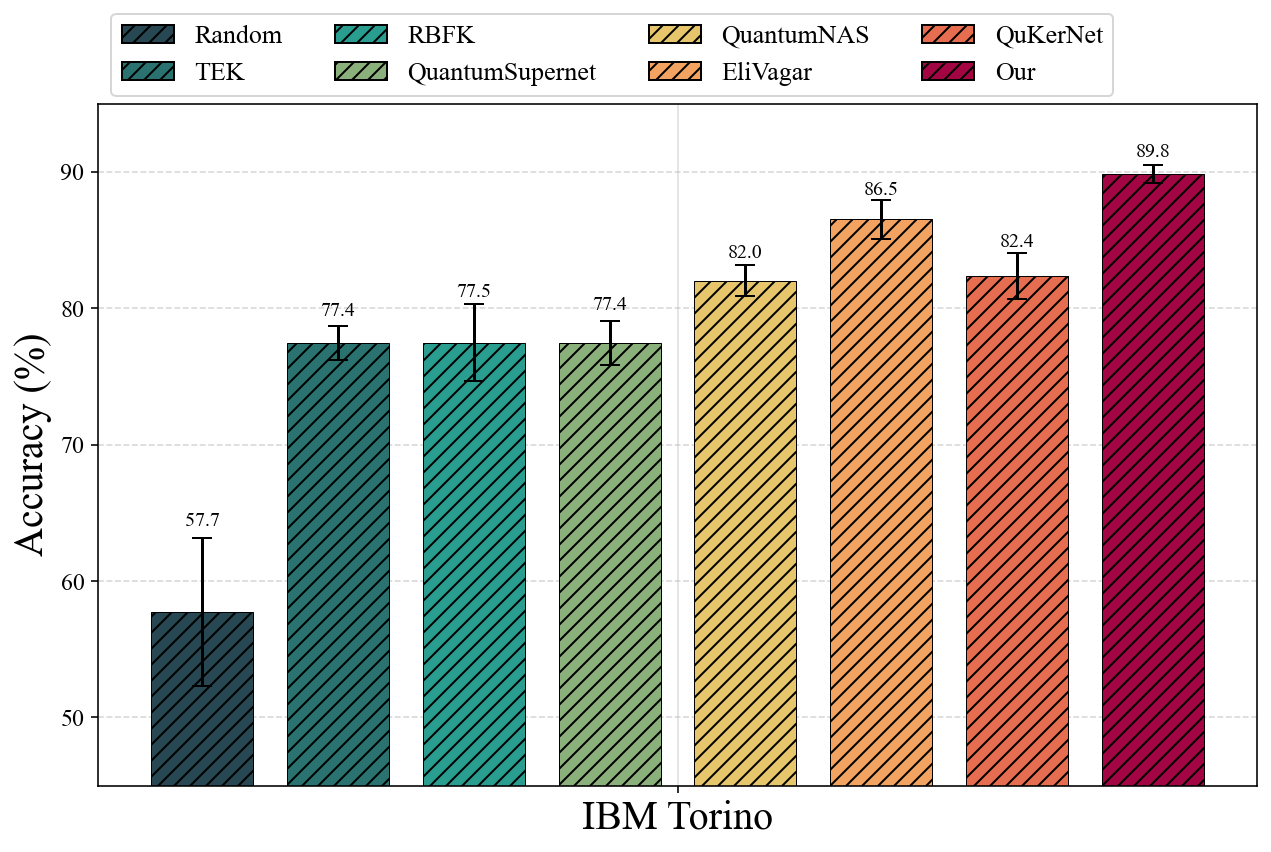}
	\caption{Classification accuracy of different methods on the FMNIST-4 classification task under noisy simulation.} \label{figureA_c}
\end{figure}

For the FMNIST-4 task, which requires 8 qubits, experiments are conducted on IBM Torino. As shown in Fig.~\ref{figureA_c}, the proposed scheme continues to achieve significantly higher classification accuracy compared to competing methods, further demonstrating its effectiveness in larger and more complex settings.

\section{Conclusion}\label{Conclusion}
In this work, we proposed a hardware-aware framework for quantum kernel design that leverages graph neural networks to evaluate and select task-specific quantum circuits under the constraints of NISQ hardware. By incorporating real-device topology, noise characteristics, and feature selection into the design pipeline, the proposed approach enables scalable and efficient kernel construction tailored to near-term quantum devices.

The use of dual graph neural network models allows for fast and accurate prediction of circuit fidelity and kernel performance, significantly reducing the computational cost of large-scale circuit screening. Experimental results on three classification benchmarks—Credit Card (CC), MNIST-5, and FMNIST-4—demonstrate superior performance compared to state-of-the-art quantum kernel methods in terms of classification accuracy. These results underscore the effectiveness of combining learning-based strategies with hardware-aware design to bridge the gap between theoretical quantum kernel models and their practical deployment on noisy quantum devices.

Future work will explore the transferability and generalization of learned quantum kernels across different tasks and data distributions, with the goal of improving their reusability in broader quantum machine learning applications. In addition, extending the framework to larger-scale quantum systems beyond the reach of classical simulation remains an important direction. This may be achieved through scalable approximation techniques, such as tensor-network-based simulations, or through direct evaluation on real quantum hardware in non-simulable regimes.


\medskip
\textbf{Abbreviations}
QML, Quantum Machine Learning; NISQ, Noisy Intermediate-Scale Quantum; GNNs, Graph Neural Networks; DAG, Directed Acyclic Graph; PST, Probability of Successful Trials; KTA, Kernel-Target Alignment; CC, Credit Card; NQE, Neural Quantum Embedding; SVMs, Support Vector Machines; RBFK, Radial Basis Function Kernel; RBF, Radial Basis Function.

\medskip
\textbf{Data Availability Statement} \par
No/Not applicable (this manuscript does not report data generation or analysis).

\section{Declarations}

\medskip
\subsection{Ethical Approval and Consent to participate}
Not applicable.

\medskip
\subsection{Consent for publication}
All authors have approved the publication. The research in this work did not involve any human, animal, or other
participants.

\medskip
\subsection{Funding}
This work is supported by the Jiangsu Funding Program for Excellent Postdoctoral Talent No.2022ZB139, the Natural Science Foundation of Jiangsu Province Higher Education Basic Science No. 24KJB120005, the Innovation Program for Quantum Science and Technology No. 2021ZD0302901, National Natural Science Foundation of China No. 62471126, Frontier Technologies R\&D Program of Jiangsu No. BF2025066, the Jiangsu Key R\&D Program Project No. BE2023011-2, the Fundamental Research Funds for the Central Universities No. 2242022k60001, and Open Project of the State Key Laboratory of Millimeter Waves No.KN202502-09.

\medskip
\subsection{Competing interests}
The authors declare no conflict of interest.

\begin{appendices}

\section{Quantum gate matrix representations}
\label{Appendix_Add}
We use the gate sets $S=\left \{ R_{x}, R_{z}, CZ, I \right \} $ and $S^{\ast } =\left \{ R_{z}, X, CNOT, I \right \}$, which primarily consist of the 1-qubit gates $I$, $X$, $R_{x}$ and $R_{z}$, whose matrix representations are given as follows:
\begin{align}
	I=\begin{pmatrix}
		1&&0 \\
		0&&1
	\end{pmatrix}
	&& X =\begin{pmatrix}
		0&& 1\\
		1&&0
	\end{pmatrix}
\end{align}
\begin{align}
	R_{x} \left ( \theta  \right ) =\begin{pmatrix}
		\cos\frac{\theta}{2}  & -i\sin\frac{\theta}{2} \\
		-i\sin\frac{\theta}{2} & \cos\frac{\theta}{2}
	\end{pmatrix}
	&& R_{z} \left ( \theta  \right ) =\begin{pmatrix}
		e^{-i\frac{\theta}{2}}  & 0 \\
		0 & e^{i\frac{\theta}{2}}
	\end{pmatrix}
\end{align}
In addition, the 2-qubit gates $CNOT$ and $CZ$ are also included, with their corresponding matrix representations defined as:
\begin{align}
	CNOT=\begin{pmatrix}
		1 && 0 && 0 && 0\\
		0 && 1 && 0 && 0\\
		0 && 0 && 0 && 1\\
		0 && 0 && 1 && 0
	\end{pmatrix}
	&& CZ=\begin{pmatrix}
		1 && 0 && 0 & 0\\
		0 && 1 && 0 & 0\\
		0 && 0 && 1 & 0\\
		0 && 0 && 0 & -1
	\end{pmatrix}
\end{align}

\section{Strategy for selecting subgraphs from the device topology}
\label{Appendix_A}

\begin{figure*}[htbp]
  \centering
  \begin{subfigure}{0.48\textwidth}
    \centering
    \includegraphics[width=\linewidth]{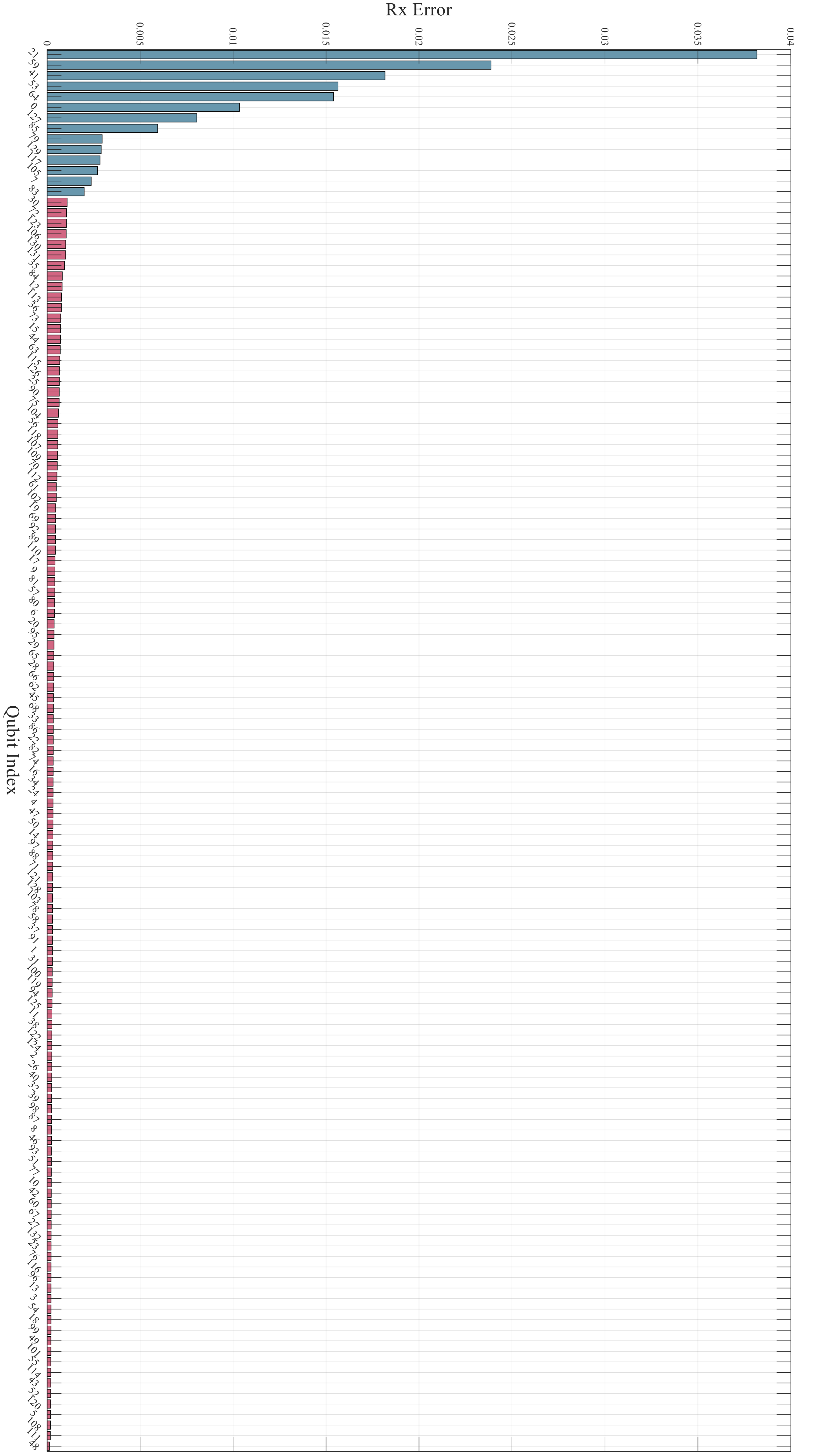}
    \caption{$R_{x}$ gate errors of different qubits on IBM Torino (sorted in descending order)}
    \label{fig:rx_errors}
  \end{subfigure}%
  \hfill
  \begin{subfigure}{0.48\textwidth}
    \centering
    \includegraphics[width=\linewidth]{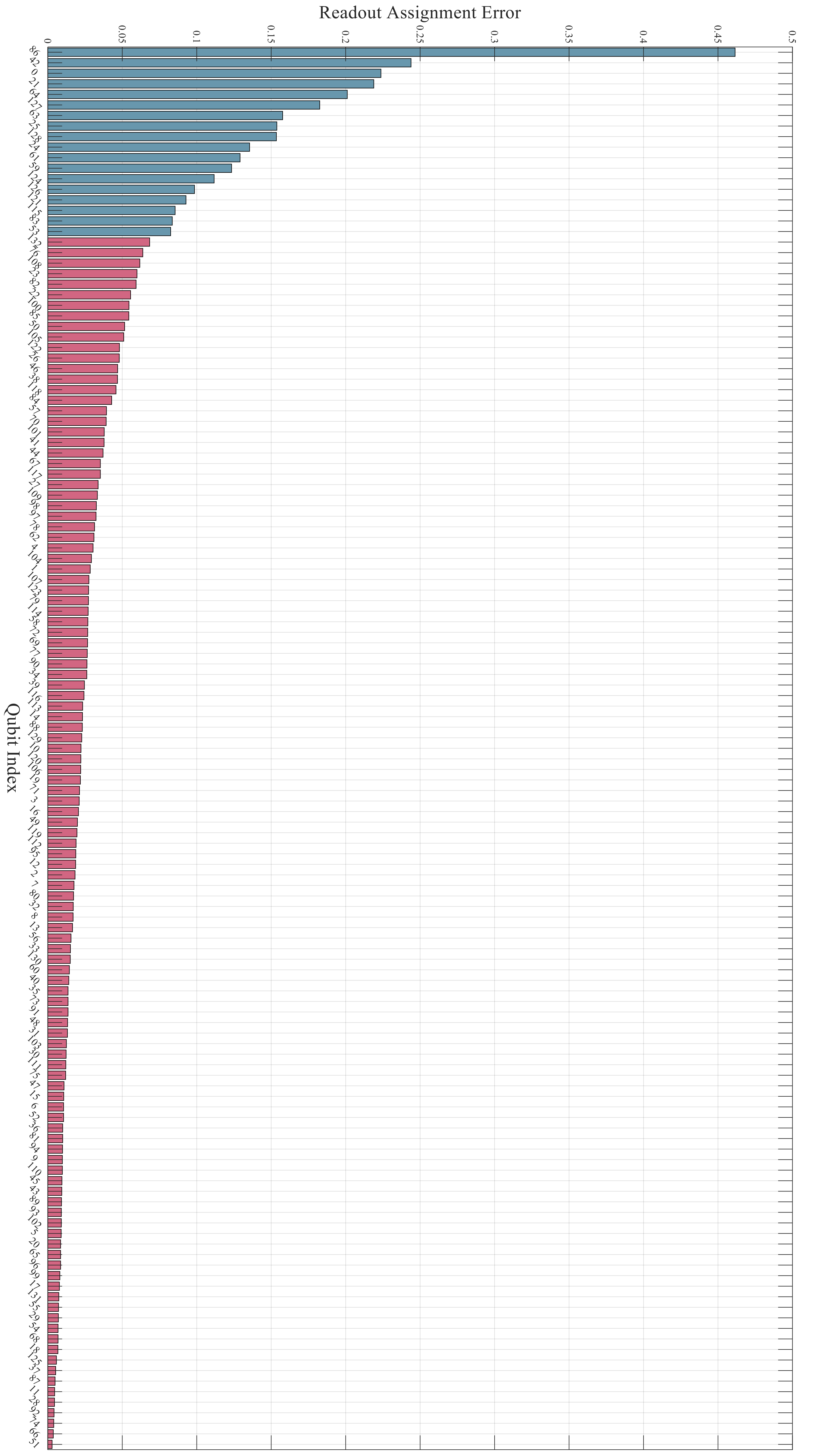}
    \caption{Readout assignment errors of different qubits on IBM Torino (sorted in descending order)}
    \label{fig:readout_errors}
  \end{subfigure}
  
  \caption{Noise characteristics of different qubits on IBM Torino.}
  \label{figureA_12}
\end{figure*}

In our investigation of quantum device noise characteristics, we observed a rapid decay in certain types of errors for large-scale quantum devices, such as the 133-qubit IBM Torino. Specifically, the 1-qubit gate error, 2-qubit gate error, and readout assignment error exhibit a sharp decline when ranked across qubits. As illustrated in Fig. \ref{figureA_12}, (a) presents the $R_{x} $ gate errors sorted in descending order, while (b) shows the readout assignment errors also sorted from highest to lowest.

\begin{algorithm}
\caption{Select subgraph from the device topology} \label{alg2}
\begin{algorithmic}[1]
	\Require 
	Device topology $G^{\ast }\left ( \left \{ V_{i}^{\ast }   \right \}, \left \{ E_{j}^{\ast } \right \}   \right )$; Excluded value $E_{xc}$; The number of qubits $N$; Noise types $T_{noise} $.
	\Ensure
	Subgraph $G$.
	\For{$type$ in range($T_{noise} $)}
	\If{$type$ related to a single qubit}
	\State Sort all qubits by error in descending order and select the top $E_{xc}$ qubits with the highest error, denoting their associated vertices and edges as $\left \{ V ^{\ast }\right \}$ and $\left \{ E ^{\ast }\right \} $.
	\State $G^{\ast }= G^{\ast }\setminus \left ( \left \{ V^{\ast } \right \},\left \{ E^{\ast } \right \}   \right ) $ \Comment{Remove $\left \{ V^{\ast } \right \}$ and $\left \{ E^{\ast } \right \} $ from topology $G^{\ast }$.}
	\ElsIf{$type$ related to two qubits}
	\State Sort all connections by error in descending order and select the top $E_{xc}$ connections with the highest error, denoting their associated edges as $\left \{ E^{\ast } \right \} $.
	\State $G^{\ast }=G^{\ast }\setminus \left \{ E^{\ast } \right \} $ \Comment{Remove $\left \{ E^{\ast } \right \} $ from topology $G^{\ast }$.}
	\EndIf
	\EndFor
	\State Select a subgraph $G$ from the topology $G^{\ast }$ with $N$ qubits and high degree of connectivity. \Comment{The degree of connectivity is defined as the maximum vertex degree.}
	\State
	\Return Subgraph $G$ 
\end{algorithmic}
\end{algorithm}

From Fig. \ref{figureA_12} (a), we observe that after excluding the top 14 qubits with the highest $R_x$ gate errors (highlighted by blue bars), the remaining 119 qubits exhibit relatively uniform $R_x$ errors. A similar trend is observed in Fig. \ref{figureA_12} (b): after excluding the top 18 qubits with the highest readout errors, the error differences among the remaining qubits become much smaller. Based on these observations, we propose a subgraph selection algorithm to identify qubits with more consistent noise characteristics. The detailed procedure is presented in Algorithm \ref{alg2}. The main idea of Algorithm \ref{alg2} is to remove vertices (qubits) and connecting edges that are heavily affected by noise from the device topology, resulting in a subgraph with low noise impact.

\section{Qubit scalability of GNN-1 and GNN-2}
\label{Appendix_B}
\begin{figure}[htbp]
    \centering
    \includegraphics[width= 0.75 \textwidth]{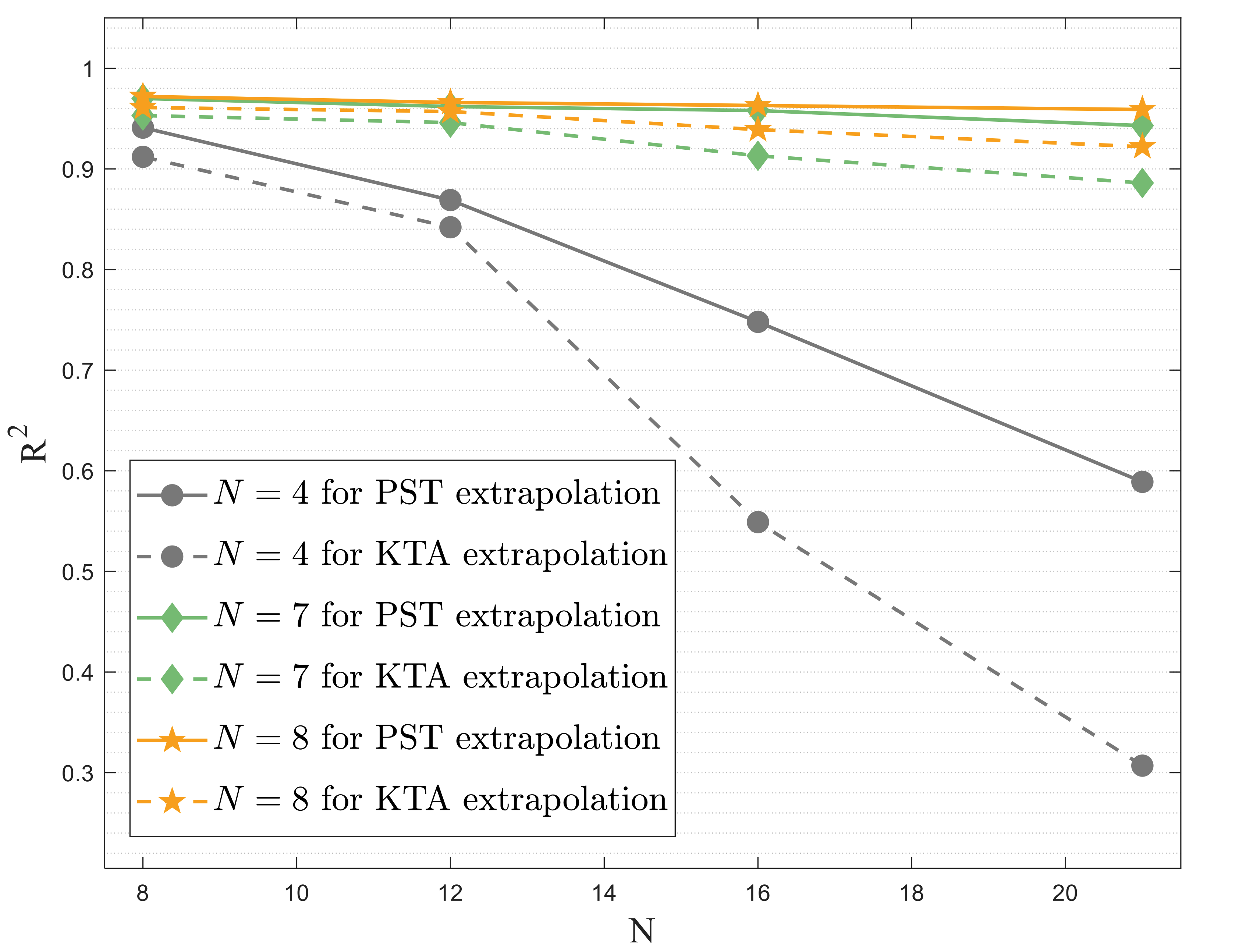}
    \caption{Qubit scalability of GNN-1 and GNN-2 for PST and KTA prediction.} \label{figureA_b}
\end{figure}

To investigate the qubit scalability of GNN-1 and GNN-2, we conducted a simulation experiment based on the IBM Torino device, which features 133 qubits. Specifically, we extended the “Target Qubit” component of the node feature vectors used in Fig. \ref{figure3_d} from the dimension of 16 to 21 to accommodate larger subgraphs. Fig. \ref{figureA_b} presents the scalability performance of GNN-1 for predicting PST and GNN-2 for predicting KTA across varying qubit numbers.
Overall, we observe that as the number of qubits increases, models trained on smaller subgraphs experience notable performance degradation when applied to circuits with more qubits. In particular, models trained on $N = 4$ qubits show a sharp decline in predictive accuracy as qubit count increases. In contrast, models trained on larger subgraphs (e.g., $N = 7$ or $8$) exhibit only a gradual decline in performance, suggesting improved scalability. These findings imply the existence of a threshold in model generalizability across qubit counts, aligning with the results reported in \cite{cantori2023supervised}, where convolutional neural networks (CNNs) were used to predict expectation values of quantum circuits.

\end{appendices}

\clearpage
\bibliography{apssamp}

\end{document}